\documentclass{aa}
\usepackage{natbib}
\bibpunct{(}{)}{;}{a}{}{,} 

\usepackage[english]{babel}           
\usepackage{graphicx,epsfig}
\usepackage{txfonts}
\usepackage{mathrsfs}
\usepackage{lscape}
\usepackage{eucal}
\usepackage{pifont}
\usepackage{rotating}
\usepackage{multirow}
\usepackage{tikz}
\usepackage{hyperref}
\usepackage{longtable}
\usepackage{wasysym}

\def \cc    {\ifmmode{\,{\rm cm}^{-3}}\else{$\,{\rm cm}^{-3}$}\fi}
\def \cq    {\ifmmode{\,{\rm cm}^{-2}}\else{$\,{\rm cm}^{-2}$}\fi}
\def \mic   {\ifmmode{\,\mu{\rm m}}\else{$\mu$m}\fi}
\def \eccs  {\ifmmode{\,{\rm erg}\,{\rm cm}^{-3} {\rm s}^{-1}}\else{$\,{\rm erg}\,{\rm cm}^{-3} {\rm s}^{-1}$}\fi}
\def \ecc   {\ifmmode{\,{\rm erg}\,{\rm cm}^{-3}}\else{$\,{\rm erg}\,{\rm cm}^{-3}$}\fi}
\def \ecm   {\ifmmode{\,{\rm erg}\,{\rm cm}^{-1}}\else{$\,{\rm erg}\,{\rm cm}^{-1}$}\fi}
\def \ecqs  {\ifmmode{\,{\rm erg}\,{\rm cm}^{-2}\,{\rm s}^{-1}\,{\rm 
             sr}^{-1}}\else{$\,{\rm erg}\,{\rm cm}^{-2}\,{\rm s}^{-1}\,{\rm sr}^{-1}$}\fi}
\def \deg   {\ifmmode{^{\circ}}\else{$^{\circ}$}\fi} 
\def \pc    {\ifmmode{\,{\rm pc}}\else{$\,{\rm pc}$}\fi} 
\def \kms   {\ifmmode{\,{\rm km}\,{\rm s}^{-1}}\else{km s$^{-1}$}\fi} 
\def \kmspc {\ifmmode{\,{\rm km}\,{\rm s}^{-1}\,{\rm pc}^{-1}}\else{km s$^{-1}$ pc$^{-1}$}\fi} 
\def \MJysr {\ifmmode{\,{\rm MJy\,sr}^{-1}}\else{$\,{\rm MJy\,sr}^{-1}$}\fi} 
\def \Kkms  {\ifmmode{\,{\rm K\,km\,s}^{-1}}\else{$\,{\rm K\,km\,s}^{-1}$}\fi}
\def \epso{\ifmmode{\overline{\varepsilon}}\else{$\overline{\varepsilon}$}\fi}
\def \utM{\ifmmode{u_{\theta,{\rm M}}}\else{$u_{\theta,{\rm M}}$}\fi}
\def \urM{\ifmmode{u_{r,{\rm M}}}\else{$u_{r,{\rm M}}$}\fi}
\def \zetaHH{\ifmmode{\zeta_{\HH}}\else{$\zeta_{\HH}$}\fi}
\def \zetaH {\ifmmode{\zeta_{\rm H}}\else{$\zeta_{\rm H}$}\fi}
\def \twCO{\ifmmode{\rm ^{12}CO}\else{$\rm^{12}CO$}\fi} 
\def \thCO{\ifmmode{\rm ^{13}CO}\else{$\rm^{13}CO$}\fi} 
\def \twCN{\ifmmode{\rm ^{12}CN}\else{$\rm^{12}CN$}\fi} 
\def \thCN{\ifmmode{\rm ^{13}CN}\else{$\rm^{13}CN$}\fi} 
\def \HdCO{\ifmmode{\rm H_{2}CO}\else{$\rm H_{2}CO$}\fi} 
\def \twHdCO{\ifmmode{\rm ^{12}H_{2}CO}\else{$\rm^{12}H_{2}CO$}\fi} 
\def \thHdCO{\ifmmode{\rm ^{13}H_{2}CO}\else{$\rm^{13}H_{2}CO$}\fi} 
\def \twC{\ifmmode{\rm ^{12}C}\else{$\rm^{12}C$}\fi} 
\def \thC{\ifmmode{\rm ^{13}C}\else{$\rm^{13}C$}\fi} 
\def \Hp{\ifmmode{\rm H^+}\else{$\rm H^+$}\fi} 
\def \Cp{\ifmmode{\rm C^+}\else{$\rm C^+$}\fi} 
\def \Sp{\ifmmode{\rm S^+}\else{$\rm S^+$}\fi} 
\def \Op{\ifmmode{\rm O^+}\else{$\rm O^+$}\fi} 
\def \CFp{\ifmmode{\rm CF^+}\else{$\rm CF^+$}\fi}
\def \CHp{\ifmmode{\rm CH^+}\else{$\rm CH^+$}\fi}
\def \CHdp{\ifmmode{\rm CH_2^+}\else{$\rm CH_2^+$}\fi}
\def \CHtp{\ifmmode{\rm CH_3^+}\else{$\rm CH_3^+$}\fi} 
\def \SHp{\ifmmode{\rm SH^+}\else{$\rm SH^+$}\fi}
\def \SHdp{\ifmmode{\rm SH_2^+}\else{$\rm SH_2^+$}\fi}
\def \SHtp{\ifmmode{\rm SH_3^+}\else{$\rm SH_3^+$}\fi}
\def \twCHp{\ifmmode{\rm ^{12}CH^+}\else{$\rm^{12}CH^+$}\fi}
\def \thCHp{\ifmmode{\rm ^{13}CH^+}\else{$\rm^{13}CH^+$}\fi}
\def \CtH{\ifmmode{\rm C_2H}\else{$\rm C_2H$}\fi} 
\def \CthHt{\ifmmode{\rm C_3H_2}\else{$\rm C_3H_2$}\fi} 
\def \Htp{\ifmmode{\rm H_3^+}\else{$\rm H_3^+$}\fi} 
\def \COp{\ifmmode{\rm CO^+}\else{$\rm CO^+$}\fi} 
\def \HCOp{\ifmmode{\rm HCO^+}\else{$\rm HCO^+$}\fi} 
\def \HtOp{\ifmmode{\rm H_3O^+}\else{$\rm H_3O^+$}\fi} 
\def \HCfiN{\ifmmode{\rm HC_5N}\else{$\rm HC_5N$}\fi} 
\def \wat{\ifmmode{\rm H_2O}\else{$\rm H_2O$}\fi} 
\def \HdO{\ifmmode{\rm H_2O}\else{$\rm H_2O$}\fi} 
\def \OHp{\ifmmode{\rm OH^+}\else{$\rm OH^+$}\fi} 
\def \HdOp{\ifmmode{\rm H_2O^+}\else{$\rm H_2O^+$}\fi} 
\def \HtOp{\ifmmode{\rm H_3O^+}\else{$\rm H_3O^+$}\fi} 
\def \NHd{\ifmmode{\rm NH_2}\else{$\rm NH_2$}\fi} 
\def \NHtrois{\ifmmode{\rm NH_3}\else{$\rm NH_3$}\fi} 
\def \oxy{\ifmmode{\rm O_2}\else{$\rm O_2$}\fi} 
\def \HH{\ifmmode{\rm H_2}\else{$\rm H_2$}\fi}
\def \Jone{\ifmmode{\rm {(J=1--0)}}\else{{(J=1--0)}}\fi} 
\def \Jtwo{\ifmmode{\rm {(J=2--1)}}\else{{(J=2--1)}}\fi} 
\def \Jthr{\ifmmode{\rm {(J=3--2)}}\else{{(J=3--2)}}\fi} 
\def \Jfou{\ifmmode{\rm {(J=4--3)}}\else{{(J=4--3)}}\fi} 
\def \Jfiv{\ifmmode{\rm {J=4--3}}\else{{J=4--3}}\fi} 
\def \Ta{\ifmmode{\rm T_A}\else{$\rm T_A$}\fi} 
\def \Tas{\ifmmode{\rm T_A^*}\else{$\rm T_A^*$}\fi} 
\def \Tmb{\ifmmode{\rm T_{mb}}\else{$\rm T_{mb}$}\fi} 
\def \Tr{\ifmmode{\rm T_r}\else{$\rm T_r$}\fi} 
\def \Trs{\ifmmode{\rm T_r^*}\else{$\rm T_r^*$}\fi}
\def \NHt{\ifmmode{N_{\rm H}}\else{$N_{\rm H}$}\fi}
\def \NH{\ifmmode{N({\rm H})}\else{$N({\rm H})$}\fi}
\def \NH2{\ifmmode{N({\rm H}_2)}\else{$N({\rm H}_2)$}\fi}
\def \NCH{\ifmmode{N({\rm CH})}\else{$N({\rm CH})$}\fi}
\def \NHF{\ifmmode{N({\rm HF})}\else{$N({\rm HF})$}\fi}
\def \dens{\ifmmode{n_{\rm H}}\else{$n_{\rm H}$}\fi}
\def \nCO{\ifmmode{n({\rm CO})}\else{$n({\rm CO})$}\fi}
\def \nHF{\ifmmode{n({\rm HF})}\else{$n({\rm HF})$}\fi}
\def \nH2{\ifmmode{n({\rm H}_2)}\else{$n({\rm H}_2)$}\fi}

\begin{document}

\title{Chemical probes of turbulence in the diffuse medium: \\ the TDR model}

\author{
  B. Godard              \inst{1},
  E. Falgarone           \inst{1}, \and
  G. Pineau des For\^ets \inst{2,1}
}

\institute{
  LERMA, CNRS UMR 8112, \'Ecole Normale Sup\'erieure \& Observatoire de Paris, Paris, France
  \and
  Institut d'Astrophysique Spatiale, CNRS UMR 8617, Universit\'e Paris-Sud, Orsay, France
  }

 \date{Received 28 January 2014 / Accepted 19 July 2014}

\abstract{
Tens of light hydrides and small molecules have now been detected over several hundreds 
sightlines sampling the diffuse interstellar medium (ISM) in both the Solar neighbourhood and 
the inner Galactic disk.}
{These new data confirm the limitations of the traditional chemical pathways driven by the
UV photons and the cosmic rays (CR) and the need for additional energy sources, such as 
turbulent dissipation, to open highly endoenergetic formation routes. 
The goal of the present paper is to further investigate the link between specific species 
and the properties of the turbulent cascade in particular its space-time intermittency.}
{We have analysed ten different atomic and molecular species in the framework of the updated 
model of turbulent dissipation regions (TDR). We study the influence on the abundances of 
these species of parameters specific to chemistry (density, UV field, and CR ionisation rate) 
and those linked to turbulence (the average turbulent dissipation rate, the dissipation 
timescale, and the ion-neutral velocity drift in the regions of dissipation).}
{The most sensitive tracers of turbulent dissipation are the abundances of \CHp\ and \SHp, 
and the column densities of the $J = 3,4,5$ rotational levels of \HH.  The abundances of 
CO, \HCOp, and the intensity of the 158 $\mu$m [CII] emission line are significantly 
enhanced by turbulent dissipation.
The vast diversity of chemical pathways allows the independent determinations of free 
parameters never estimated before: an upper limit to the average turbulent dissipation rate,  
$\epso \lesssim 10^{-23}$ \eccs\ for \dens=20 \cc, from the \CHp\ abundance; an upper limit 
to the ion-neutral velocity drift, $\upsilon_{\rm in} \lesssim 3.5$ \kms, from the \SHp\ to 
\CHp\ abundance ratio; and a range of dissipation timescales,  $100 \lesssim \tau_V \lesssim 
1000$ yr, from the CO to \HCOp\ abundance ratio. For the first time, we reproduce the large 
abundances of CO observed on diffuse lines of sight, and we show that CO may be abundant even 
in regions with UV-shieldings as low as $5\times 10^{-3}$ mag. The best range of parameters 
also reproduces the abundance ratios of OH, \CtH, and \HdO\ to \HCOp\ and are consistent 
with the known properties of the turbulent cascade in the Galactic diffuse ISM.}
{Our results disclose an unexpected link between the dissipation of turbulence and the 
emergence of molecular richness in the diffuse ISM. Some species, such as \CHp\ or \SHp, 
turn out to be unique tracers of the energy trail in the ISM. In spite of some degeneracy, 
the properties of the turbulent cascade, down to dissipation, can be captured through 
specific molecular abundances.}


\authorrunning{B. Godard, E. Falgarone, and G. Pineau des For\^ets}
\titlerunning{Chemical probes of turbulence in the diffuse ISM.}
\maketitle

\section{Introduction}

Turbulence in galaxies stands at the crossroad of a wide variety of cosmic processes: star 
and planet formation, galatic dynamos, magnetic reconnection, to name a few. It has a strong 
interplay, not only with gravity and magnetic fields, but also with the microphysics of gas 
because the scales at which it dissipates are close to the mean free path of atoms and 
molecules. The turbulent cascade in galaxies therefore channels huge amounts of supra-thermal 
kinetic energy from the large scales at which it is fed to those at which atoms and molecules 
interact.

This property of the turbulent cascade has started to be exploited to elucidate the 
long-standing puzzle raised by the large abundances of many molecular species that have 
been detected in the diffuse medium: \CHp\ \citep{Crane1995,Gredel1997,Weselak2008,
Falgarone2010,Godard2012}, \HCOp \citep{Lucas1996,Godard2010}, CO \citep{Sheffer2008,
Liszt2010}, SH and H$_2$S \citep{Neufeld2012}, \SHp\ \citep{Godard2012}, CN, HCN, and 
HNC \citep{Liszt2001,Godard2010}. Indeed these abundances cannot be understood in the 
framework of UV-driven chemistry because their formation proceeds through highly 
endoenergetic reactions that cannot be activated at the low temperature of the diffuse 
interstellar medium (ISM). These findings are not specific to the Milky Way. Large 
abundances of \CHp\ are now observed in the diffuse medium of external galaxies where 
the \CHp(1-0) line is observed in absorption against the bright dust continuum emission 
of nearby starbursts: Arp 220 \citep{Rangwala2011}; Mrk231 \citep{van-der-Werf2010}; and 
M82, Circinus, and NGC4945 (Falgarone et al., in prep.).

It has been proposed that the bursts of turbulent dissipation (i.e. its property of 
space-time intermittency) sufficiently heat the diffuse gas locally to produce the 
observed abundances \citep{Joulain1998}. This would explain why the diffuse ISM is, 
occasionally, such a strong emitter in the pure rotational lines of \HH\ \citep{
Falgarone2005,Ingalls2011}.
Godard et al. (2009, hereafter Paper I) have improved the model proposed by \citet{
Joulain1998} and set up global models of random sight lines across the medium that 
sample {\it (i)} the active dissipative regions driving a non-equilibrium warm 
chemistry, {\it (ii)} the regions in chemical and thermal relaxation that develop 
after the extinction of the dissipation bursts, and {\it (iii)} the ambient phase 
characterised by UV-driven chemistry. Although still idealised in some respects, these 
models take the energetic constraints provided by the turbulent cascade into account. 
Interestingly, it is found that less than a few percent of the turbulent energy, on 
average, is sufficient to drive the warm chemistry at the observed level, the rest 
being radiated away mostly in the pure rotational lines of \HH\ and [CII] fine-structure 
line emission.  Moreover, the comparison of the model predictions to observations 
clearly favours a turbulent dissipation dominated by ion-neutral friction in structures 
as small as $\sim$ 100 au.

In this paper, we further explore the models of Paper I and take advantage of the 
impressive amount of new data obtained on light molecular hydrides in the diffuse 
medium with {\it Herschel}. We compare the predictions of the TDR model to these
observations, and to older data sets on CO and \HCOp. 
An overview of the information directly inferred from the absorption spectroscopy data 
is given in Sect. \ref{sec-data}. The main characteristics of the model are summarised 
in Sect. \ref{Sec-TDR}. In Sects. \ref{Sect-Diagnostics} to \ref{Sect-Cp}, we explore 
the parameter space over large domains and use the model predictions to identify molecular 
species (and pairs of species) that can be used as specific diagnostics of the turbulence 
in the diffuse ISM. Finally, in Sect. \ref{Sect-discussion}, we discuss the overall 
coherence of the parameters of the turbulence inferred from the light hydrid and CO 
molecular abundances. The recent refinements of the dynamical and chemical prescriptions 
of the TDR model are given in the Appendices.

\section{Observational constraints on the parameter space to explore}\label{sec-data}

\subsection{Gas density}
The medium discussed in the present paper is the cold phase of the neutral ISM, the cold 
neutral medium (CNM). The thermal pressure of this medium has been measured with observations 
of the [CI] line by \citet{Jenkins2011} to have a mean value $P_{th}/k_B=3800$ K \cc, with 
most of the gas in the range $2\times 10^3$ to $10^4$ K\cc. The lower temperature of the 
CNM is reliably determined from HI absorption measurements $T_K \gtrsim 80$K, from which we 
infer a range of upper limits for the gas densities $n_{\rm H} \lesssim 25$ \cc\ to 125 
\cc.  \citet{Goldsmith2013} has computed the excitation temperatures of CO molecules seen 
through several absorption lines against nearby stars \citep{Sheffer2008}. He infers median 
values of the density \dens= 92 \cc\ and 68 \cc\ for gas temperatures 50 and 100 K, respectively, 
and therefore thermal pressures well within the range of Jenkins and Tripp (2011). An independent 
determination is based on [CII] and [CI] observations of the same sightlines as those where 
the \CHp(1-0) line has been detected \citep{Gerin2014}.  Self-absorption of 
the [CII] line occurs over the same velocity range as that of the \CHp(1-0) absorption (see 
also \citealt{Falgarone2010a}). The excitation analysis of the [CII] and [CI] lines at these 
velocities shows that the gas causing the \CHp(1-0) absorption is indeed the CNM, with a 
narrow range of densities from $n_{\rm H}= 50$ \cc\ to less than 200 \cc.

\subsection{Gas column densities}
For reasons linked to their excitation conditions, most of the molecular lines observed in 
the diffuse ISM are seen in absorption against continuum sources that are either quasars 
and free-free emission of HII regions, at millimetre frequencies or warm dust thermal
emission in star forming regions, at submillimetre frequencies. One exception is the CO(1-0) 
line which can be excited in gas densities as low as 50~\cc\ and provide a detectable level 
of line emission \citep{Liszt2012}. The parameters directly inferred from observations are 
the line optical depth at a given velocity and the velocity distribution.

Absorption lines corresponding to long pathlengths across the Galaxy have broad line profiles 
that are usually decomposed into individual Gaussian components. The Gaussian shape, the 
number of components, and their width are somewhat arbitrary. This is not a major limitation, 
because the total hydrogen column density (atomic and molecular) with which those of molecular 
species are associated are computed over the same velocity components found in the profiles of 
HI for the atomic gas, and CH and HF for \HH\ (e.g. \citealt{Godard2012}). The column densities 
plotted in most of the figures of the paper are those associated with these Gaussian velocity
components. This is the reason that the plotted total column densities of hydrogen are never 
much larger than $10^{22}$\cq, in spite of the long pathlengths sampled in the Galactic plane 
by the {\it Herschel} observations. Whether or not a well-defined ``cloud'' of gas is ascribed 
to each of these components is a question that goes beyond the scope of our paper. Following 
\citet{Rachford2002}, we stress that a {\it translucent line of sight} in the diffuse ISM is
definitely different from a {\it translucent cloud} since large column densities in general 
result from the accumulation on the same line of sight of a large number of fragments, each 
of low column density, and presumably poorly shielded from the ambient UV field.  This has 
been well illustrated by \citet{Levrier2012} who show, using numerical simulations of 
magnetohydrodynamical (MHD) turbulence, that the column density peaks do not correspond to 
sightlines towards density peaks in the cube. 


\subsection{UV shielding}
The very large scatter (by several orders of magnitude) of the \HH\ fraction observed on 
sightlines with total hydrogen column densities ranging between $3 \times 10^{19}$\cq\ 
and $5 \times 10^{20}$\cq\ \citep{Gillmon2006} is also an illustration that the gas on 
such diffuse sightlines is highly fragmented and that there is not a systematic increase 
in the UV-shielding with the gas column density sampled. Molecular hydrogen fractions are 
computed in Appendix B using the Meudon PDR (photodissociation regions) code\footnote{
Version 1.4.4 available at \url{http://pdr.obspm.fr/PDRcode.html}.} \citep{Le-Petit2006}
for a range of densities and UV-shieldings. The \HH\ self-shielding is so efficient that 
the sharp transition from $f_{\HH} \sim 10^{-4}$ to 0.1 or more always occurs for very 
low UV-shieldings, $A_V \lesssim 0.02$ mag, even at a density as low as 10 \cc. This is 
well below the range of observed UV-shieldings, $0.02 \lesssim A_V \lesssim 0.3$ mag, 
that can be inferred from the above range of $N_{\rm H}$ assuming $N_{\rm H}= 1.8 \times 
10^{21} A_V$\cq. The only way to reconcile what is known on the self-shielding of \HH\ 
and the observed width of the transition region, given the low densities of the diffuse 
ISM, is that this column density range is due to tens of fragments of very low column 
density and probably low UV-shielding as well.

\subsection{The free parameters linked to turbulence in the diffuse ISM}

The CNM is a highly turbulent medium and its supra-thermal kinetic and magnetic pressures 
provide it with the required support to reach quasi-static equilibrium in the gravitational
potential of the Galaxy \citep{Cox2005}. Its weak ionisation degree, carried by \Cp\ ions, 
makes it not fully coupled to the magnetic fields. Turbulence in this cold phase of the 
interstellar medium is therefore magnetised, compressible, and multifluid, with a high
Reynolds number. The power spectra of both the velocity and the column density inferred
from HI line emission are found to have the slope of the Kolmogorov incompressible turbulence 
(e.g. in the North Celestial Loop complex, \citealt{Miville-Deschenes2003}). The energy 
transfer rate estimated in the atomic turbulent cascade, is found to be the same as that 
inferred in the diffuse molecular gas, traced by the optically thick CO(1-0) line emission, 
suggesting that diffuse molecular and atomic gas are part of the same turbulent cascade 
\citep{Hennebelle2012}. An interesting feature is the large range of fluctuations for this 
energy transfer rate: it fluctuates by two orders of magnitudes above and below the average
value, $\epso_{obs}=2 \times 10^{-25}$ \eccs, and this is independent of the scale at which 
these fluctuations are measured.
The non-thermal rms velocity dispersion at large scale (between 10 pc and 100 pc) ranges 
between about 1 \kms\ and 8.5 \kms\ from the size-linewidth relation in diffuse molecular 
gas \citep{Falgarone2009a}. The column density-weighted rms velocity of CNM HI Gaussian 
components in the Solar neighbourhood is $\sim 7$ \kms\ \citep{Heiles2003} while the 
distributions of HI linewidths for the coldest components are 4.9 \kms\ and 12 \kms\ at 
larger scale in the Galaxy \citep{Haud2007}.  These quantities may be understood as due 
to turbulence in the diffuse medium \citep{Kalberla2009,Kim2013}.

\section{Overview of the TDR model} \label{Sec-TDR}

The model of turbulent dissipation regions (TDR) is built on the property of space-time 
intermittency of turbulence, i.e. its dissipation is not evenly distributed in space but 
concentrated in bursts. The purpose of the TDR model is to predict the molecular abundances 
observed in the diffuse ISM taking into account those transiently formed in the warm 
non-equilibrium chemistry driven by the bursts of turbulent dissipation, in addition to 
those formed in the ambient medium, via UV-driven chemistry. The proper description of the 
dissipation regions of compressible MHD turbulence being far beyond 
the grasp of even the most powerful direct numerical simulations \citep{Kritsuk2007}, the 
approach followed in the model is idealised, but relies on three universal properties of 
turbulence. {\it (i) } The bulk of the kinetic energy in compressible turbulence lies in 
incompressible modes because of the fast energy transfer from compressible to solenoidal 
modes \citep{Porter2002,Vestuto2003,Kritsuk2010}. {\it (ii)} A significant fraction of the 
turbulent energy in incompressible flow experiments is dissipated in coherent dynamical 
structures, such as vortices \citep{Douady1991,Nagaoka2002, Mouri2007,Mouri2009}. Similar 
structures are found in numerical simulations \citep{Moisy2004, Kritsuk2007,Uritsky2010,
Momferratos2014}. Their characteristics are set by those of the ambient turbulence 
\citep{Jimenez1997,Tanahashi2004,Mouri2009}. {\it (iii) } Lagrangian intermittency is 
observed in laboratory flows to be even more pronounced than in Eulerian statistics and 
Lagrangian accelerations of the fluid cells are influenced by the large scale dynamics 
(see the review of \citealt{Arneodo2008}).

\begin{table}[!ht]
\begin{center}
\caption{Parameters of the TDR model. The standard value and the range of explored values
are given in Cols. 3 and 4, respectively.}
\begin{tabular}{l l c c c}
\hline
             & unit        & standard            & range                                     & ref     \\
\hline
$\dens$      & \cc         & $50$                & $20 - 300$                                & $a$     \\
$\chi$       &             & 1                   & 0.5 - 10                                  & $b$     \\
$A_V$        & mag         & 0.4                 & 0.005 - 1                                 & $c$     \\
$\zetaHH$    & s$^{-1}$    & $10^{-16}$          & $3 \times 10^{-17}$ - $3 \times 10^{-16}$ & $d,e$   \\
\hline
$a$          & s$^{-1}$    & $3 \times 10^{-11}$ & $10^{-11}$ - $10^{-10}$                   & $f$     \\
$\utM$       & km s$^{-1}$ & $3.0$               & 1 - 7                                     & $g,h,i$ \\
$\epso$      & \eccs       & $10^{-24}$          & $10^{-25}$ - $10^{-23}$                   & $j$     \\
$E_{\tau_V}$ & \ecm        & $3 \times 10^{-13}$ & $10^{-13}$ - $10^{-11}$                   &         \\
\hline
\end{tabular}
\begin{list}{}{}
($a$) \citet{Snow2006},     ($b$) \citet{Mathis1983},    ($c$) \citet{Godard2012},
($d$) \citet{Dalgarno2006}, ($e$) \citet{Indriolo2012},
($f$) \citet{Godard2009},   ($g$) \citet{Tanahashi2004}, ($h$) \citet{Mouri2007},
($i$) \citet{Mouri2009},    ($j$) \citet{Hennebelle2012}
\end{list}
\label{Tab-Par}
\end{center}
\end{table}

Numerically, the TDR model is a three-step code (Paper I). It first computes the non-equilibrium 
thermal and chemical Lagrangian evolutions of the gas during the active dissipation stage. 
Then, after the end of the dissipation burst, the thermal and chemical relaxation of the 
gas are computed as functions of time. In a third step, the column densities of all the 
chemical species treated by the code are computed along a random line of sight across the 
diffuse ISM, taking into account energetic constraints. We briefly summarise below the main 
features of each of these steps, all the details being given in Paper I. The free parameters 
of the model and the range of values explored in the present work are given in Table 
\ref{Tab-Par}.

\subsection{The active dissipative stage}

This stage is treated in a hybrid way. Following the above, we adopt a modified Burgers vortex 
\citep{Nolan1999} for the coherent structures responsible for intermittent dissipation. Such a 
vortex results from the balance between the diffusion and the stretching of the vorticity by the
rate-of-strain. The vortex is defined by two independent parameters, the maximum orthoradial 
velocity $\utM$ and the turbulent rate-of-strain $a$ (see Table \ref{Tab-Par}). The vortex 
equilibrium radius is $r_0^2=4 \nu/a$, where $\nu$ is the molecular viscosity. We first compute 
the steady-state equilibrium of such a vortex of finite length threaded by a magnetic field, 
initially parallel to the vorticity \citep{Mininni2006,Mininni2006a}. The steady-state 
configuration is reached within $\sim 100$ yr with a slightly helical field and ion velocities 
at least 10 times smaller than $\utM$.



In this configuration, the gas is mechanically heated through viscous friction and 
ion-neutral friction, at an average rate per unit volume $\overline{\Gamma}_{\rm turb}$ 
computed up to a radius $Kr_0$ ($K\sim 5$, see Paper I) where the turbulent heating stops 
influencing the gas temperature and chemistry. The gas cools down through atomic and 
molecular line emissions. The duration of this phase (i.e. the lifetime of the vortex), 
$\tau_V$, is related to the total energy dissipated per unit length of the vortex\footnote{
The vortex is invariant along its axis.} over its lifetime and volume (hereafter called 
the vortex dissipation integral), 
\begin{equation} \label{Eq-Etau}
E_{\tau_V}=\pi (Kr_0)^2\overline{\Gamma}_{\rm turb} \tau_V,
\end{equation}
a free parameter also introduced in Paper I. The non-equilibrium coupled thermal and chemical 
evolutions of the gas are computed along the Lagrangian trajectory of a fluid particle trapped 
in the vortex. The code simultaneously solves the time-dependent evolution of 11 dynamical and 
181 chemical variables, including the abundances of 163 species and of the first 18 rotational 
levels of \HH\ (see Appendix \ref{Append-Chemistry}). The rates of all collisional processes 
(including chemical reactions) are computed at an effective temperature \citep{Flower1985},
\begin{equation} \label{Eq-Teff}
T_{\rm eff} = \frac{m_1 T_2 + m_2 T_1}{m_1 + m_2} + \frac{1}{k} \frac{m_1 m_2}{m_1 + m_2} u_D^2,
\end{equation}
where $m_1$, $m_2$, $T_1$, and $T_2$ are the masses and temperatures of the two colliders,
$u_D$ is their relative velocity, and $k$ is the Boltzmann constant.

\subsection{The relaxation stage}

\begin{figure*}[!ht]
\begin{center}
\includegraphics[width=17.0cm,angle=-0]{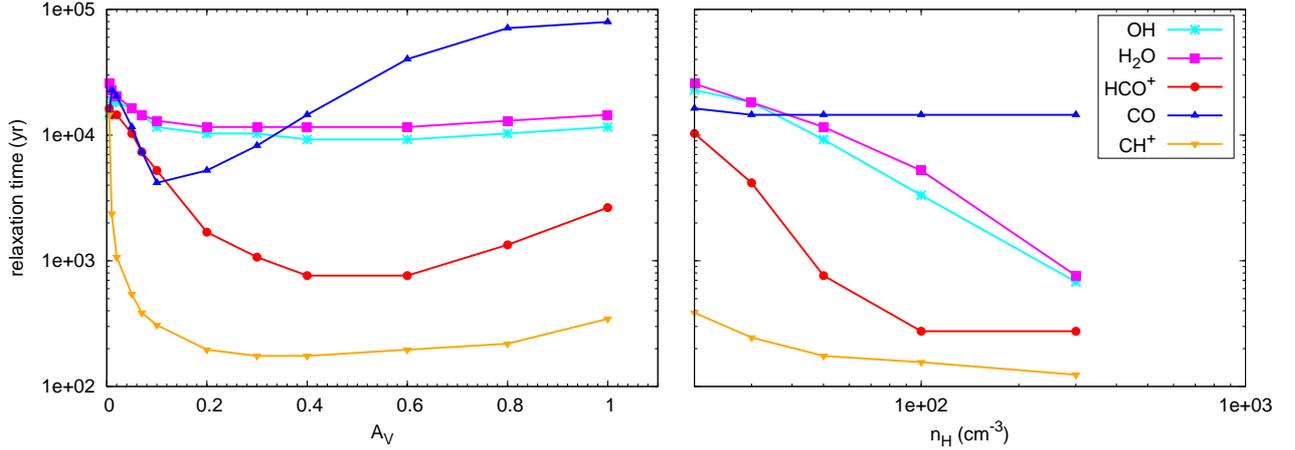}
\caption{Relaxation timescales (Eq. \ref{Eq-relax-time}) of OH (cyan crosses), \HdO\ (magenta
squares), \HCOp\ (red circles), CO (blue triangles), and \CHp\ (orange upside-down triangles) 
computed with the TDR model as functions of the shielding $A_V$ of the UV field (left panel) 
and the density $\dens$ of the gas (right panel), all the other parameters being set to their
standard values (see Table \ref{Tab-Par}).}
\label{Fig-relax-time}
\end{center}
\end{figure*}

By nature, an active dissipation burst has a short lifetime $\tau_V$. Once it ends, the gas 
cools down and the chemical signatures imprinted by the active stage persist from a few 
hundred years to a few 10$^4$ years depending on the destruction timescales of the different 
species. An Eulerian approach is adopted here: the code computes the chemical and thermal 
evolutions of 50 points evenly distributed along the radial axis of the vortex, now at rest, 
assuming that the gas obeys an isochoric or isobaric equation of state. All the results 
presented in the following are obtained for an isochoric relaxation. The column densities of 
any species X integrated across the vortex, $N_R({\rm X},t)$, are therefore computed as 
functions of time.

The relaxation timescales $\tau_R({\rm X})$ defined as
\begin{equation} \label{Eq-relax-time}
\int_0^{\tau_R({\rm X})} N_R({\rm X},t) dt = 0.9 \int_0^{\infty} N_R({\rm X},t) dt
\end{equation}
are shown in Fig. \ref{Fig-relax-time} as functions of the shielding of the UV field (left 
panel) and the density of the gas (right panel) for \CHp\ and four oxygen bearing species, 
OH, \HdO, \HCOp, and CO. This figure illustrates the very large scatter among the chemical 
relaxation timescales: from a few 100 yr for \CHp\ and \SHp\ to more than $10^4$ yr for CO. 
Among all the species treated in our chemical network (see Appendix \ref{Append-Chemistry}), 
CO has the largest relaxation timescale as long as $A_V \gtrsim 0.3$ mag and $\dens \gtrsim 
30$ \cc. This is due to the shielding of the molecule from the UV radiation field 
(see Appendix \ref{Append-H2-CO}).

\subsection{Building of a line of sight} \label{Sec-build-los}

\begin{figure}[!hb]
\begin{center}
\includegraphics[width=8.5cm,angle=0]{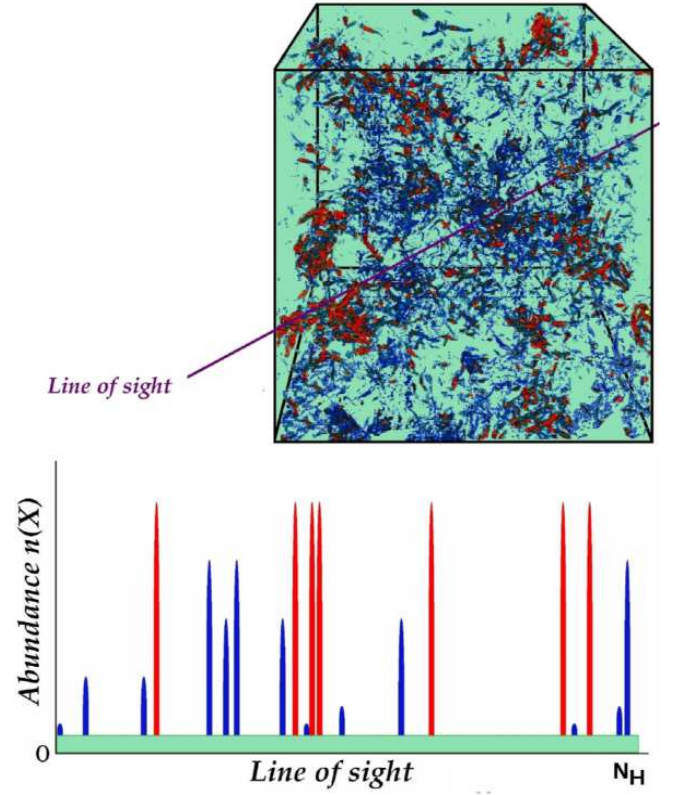}
\caption{Schematic view of the TDR model. A random line of sight across the diffuse interstellar 
medium intercepts the ambient UV-dominated gas (in green), a number of active vortices (in 
red), and the corresponding number of relaxation stages (in blue). The chemical enrichment 
(shown in the lower panel) only occurs on a small fraction of the entire line of sight
(a few percent, Paper I).}
\label{FigTDR2}
\end{center}
\end{figure}

\subsubsection{The three phases sampled by a random line of sight}

In Fig. \ref{FigTDR2} we display a schematic view of the turbulent diffuse interstellar 
medium. In this conception, most of the gas volume is filled by a non-dissipative medium 
(in green; hereafter called the ambient medium) where the chemical composition of 
the gas solely results from its density $\dens$, the strength $\chi$ of the UV radiation 
field (expressed in Mathis' unit, \citealt{Mathis1983}), the shielding $A_V$ of the ISRF 
(at visible wavelength) at each position, and the total CR ionisation rate of molecular 
hydrogen $\zetaHH$ (taking into account secondary ionisations). In addition to this 
ambient medium, a random line of sight intercepts a number 
$\mathscr{N}_V$ of active vortices (in red) and a number of relaxation stages (in blue). 
Since each relaxation phase follows an active dissipation phase, their number along the 
line of sight at any time is proportional to $\mathscr{N}_V$ and to $\tau_R({\rm X})/
\tau_V$ where $\tau_R({\rm X})$ is the relaxation timescale of species X. Introducing the 
filling factor of the active vortices on a line of sight of length $L$, 
\begin{equation}
f_L=\mathscr{N}_V \,2Kr_0/L, 
\end{equation}
the column density of a species X writes\footnote{There is an unfortunate misprint in this 
Equation in paper I.}
\begin{equation} \label{Eq-Coldens}
N({\rm X}) = N_{\rm amb}({\rm X}) + f_L N_V({\rm X}) + f_L {\tau_R({\rm X}) \over \tau_V} 
{\overline N_R({\rm X})},
\end{equation}
where $N_{\rm amb}({\rm X})$ is its column density in the ambient medium, $N_V({\rm X})$ 
its column density in an active vortex, and ${\overline N_R({\rm X})}$ its averaged column 
density during the relaxation phase. We note that according to Fig. 
\ref{FigTDR2} (lower panel), $N_V({\rm X})$ and ${\overline N_R({\rm X})}$ are computed as 
excesses\footnote{These quantities may therefore be negative.} of column densities compared 
to $N_{\rm amb}({\rm X})$. We also note that $f_L$ being defined on any random line of sight,
it corresponds to a volume filling factor. It thus relates the dissipation rate averaged at 
large scale \epso\ to the same quantity averaged over the volume of a 
vortex: 
\begin{equation}
\epso=f_L \,\overline{\Gamma}_{\rm turb}.
\end{equation}

Aside from \dens, $\chi$, $A_V$, and $\zetaHH$, the TDR model is therefore defined by four 
independent parameters: the turbulent rate-of-strain $a$; the vortex maximum orthoradial 
velocity $\utM$ (which sets the ion-neutral velocity drift $\upsilon_{in}$); 
the vortex dissipation integral $E_{\tau_V}$ (involving the vortex lifetime $\tau_V$); 
and the average dissipation rate \epso, related to the volume filling factor $f_L$ of 
all the active vortices.

\subsubsection{Prescriptions for the shielding of the UV radiation field}


The UV radiation field being a main input of the gas phase chemistry, the prescription 
for the radiative transfer is critical. In this paper, we adopt two different descriptions 
of the variation of $A_V$ along the line of sight. In the first approach (standard case), 
a homogeneous FUV shielding is adopted along the entire line of sight: the medium is 
considered as highly fragmented, each fragment being evenly shielded by its environment. 
In the second approach, we mimic PDR models: we consider a 1D slab illuminated from both 
sides with the same UV radiation field (parametrised by $\chi$) and adopt a shielding 
function that linearly increases from the edge to the mid plane of the slab and then 
linearly decreases up to the opposite edge.

In the following, the predictions of the model will always correspond to those 
obtained with the first approach, except when indicated otherwise.

\section{Chemical diagnostics of turbulent dissipation} \label{Sect-Diagnostics}

\subsection{Exploration of the parameter space} \label{Sect-explo-param}

The influence of the parameters of the TDR model on the vortex dynamics and lifetime, the 
thermal evolution of the gas, and the distributions of the active and relaxation phases 
along the line of sight have been extensively described in Paper I. In the present paper, 
we broaden the parameter domain explored and focus on their impact on the chemical composition 
of a random line of sight.

The elemental abundances are set to their values observed in the Solar neighbourhood (SN, 
see Table \ref{Tab-Ele} in Appendix \ref{Append-Chemistry}). We explore the parameter domain 
of the code through a grid of $\sim$ 20,000 models\footnote{Since $\epso$ and $E_{\tau_V}$ 
are only used for building a line of sight, there is no need to run additional models when we 
change these parameters.} (corresponding to a total computing time of $\sim 500$ hours) 
organised as follows:
\begin{itemize}
\item[$\bullet$] 5 densities $\dens$ = 20, 30, 50, 100, and 300 \cc\ to bracket the range 
discussed in Sect. \ref{sec-data};
\item[$\bullet$] 4 radiation field scaling factors $\chi$ = 0.5, 1, 3, and 10, to take into 
account variations from the SN to the inner Galactic disk (cf. Sect. \ref{Sect-Var-CHp});
\item[$\bullet$] 12 visible extinctions $A_V$ = 0.005, 0.01, 0.02, 0.05, 0.07, 0.1, 0.2, 
0.3, 0.4, 0.6, 0.8, and 1.0 mag to explore media with very low UV shielding as discussed
in Sect. \ref{sec-data}; 
\item[$\bullet$] 3 cosmic ray ionisation rates $\zetaHH = 3 \times 10^{-17}$, $10^{-16}$, 
and $3 \times 10^{-16}$ s$^{-1}$ to explore the largest average value recently derived in the 
diffuse ISM \citep{Padovani2013,Indriolo2012};
\item[$\bullet$] 3 rates-of-strain $a$ = 1, 3, and 10 $\times 10^{-11}$ s$^{-1}$ (Paper I);
\item[$\bullet$] 9 maximum orthoradial velocities $\utM$ = 1, 2, 2.5, 3, 3.5, 4, 5, 6, and 
7 \kms\ following Sect. \ref{sec-data};
\item[$\bullet$] 5 average dissipation rates $\epso$ = 0.1, 0.3, 1, 3, and 10 $\times 10^{-24}$ 
\eccs\ following Sect. \ref{sec-data}; and
\item[$\bullet$] 5 vortex dissipation integrals $E_{\tau_V}$ = 0.1, 0.3, 1, 3, and 10 
$\times 10^{-12}$ \ecm, i.e. below the value above which this parameter has no impact
(Sect. \ref{Sect-CO}) and keeping only dissipation timescales above the validity limit 
of the model (see Paper I).
\end{itemize}
In the paper, we will often refer to a standard model defined by $\dens = 50$ \cc, $\chi = 
1$, $A_V$ = 0.4 mag, $\zetaHH = 10^{-16}$ s$^{-1}$, $a = 3 \times 10^{-11}$ s$^{-1}$, $\utM 
= 3$ \kms, $\epso = 10^{-24}$ \eccs, and $E_{\tau_V} = 3 \times 10^{-13}$ \ecm. We discuss 
the influence of $\zetaHH$ on the chemistry of the gas and justify our choice for the standard
value in Sect. \ref{Sect-zeta}. Unless indicated otherwise, all the parameters are 
systematically set to their standard values.

\subsection{Method: ternary phase diagrams}

\begin{figure*}[!ht]
\begin{center}
{\bf \Large \underline{constant \dens}}
\hspace{7.5cm}
{\bf \Large \underline{constant $A_V$}} \\
\vspace{0.5cm}
\includegraphics[width=7.5cm,angle=-0]{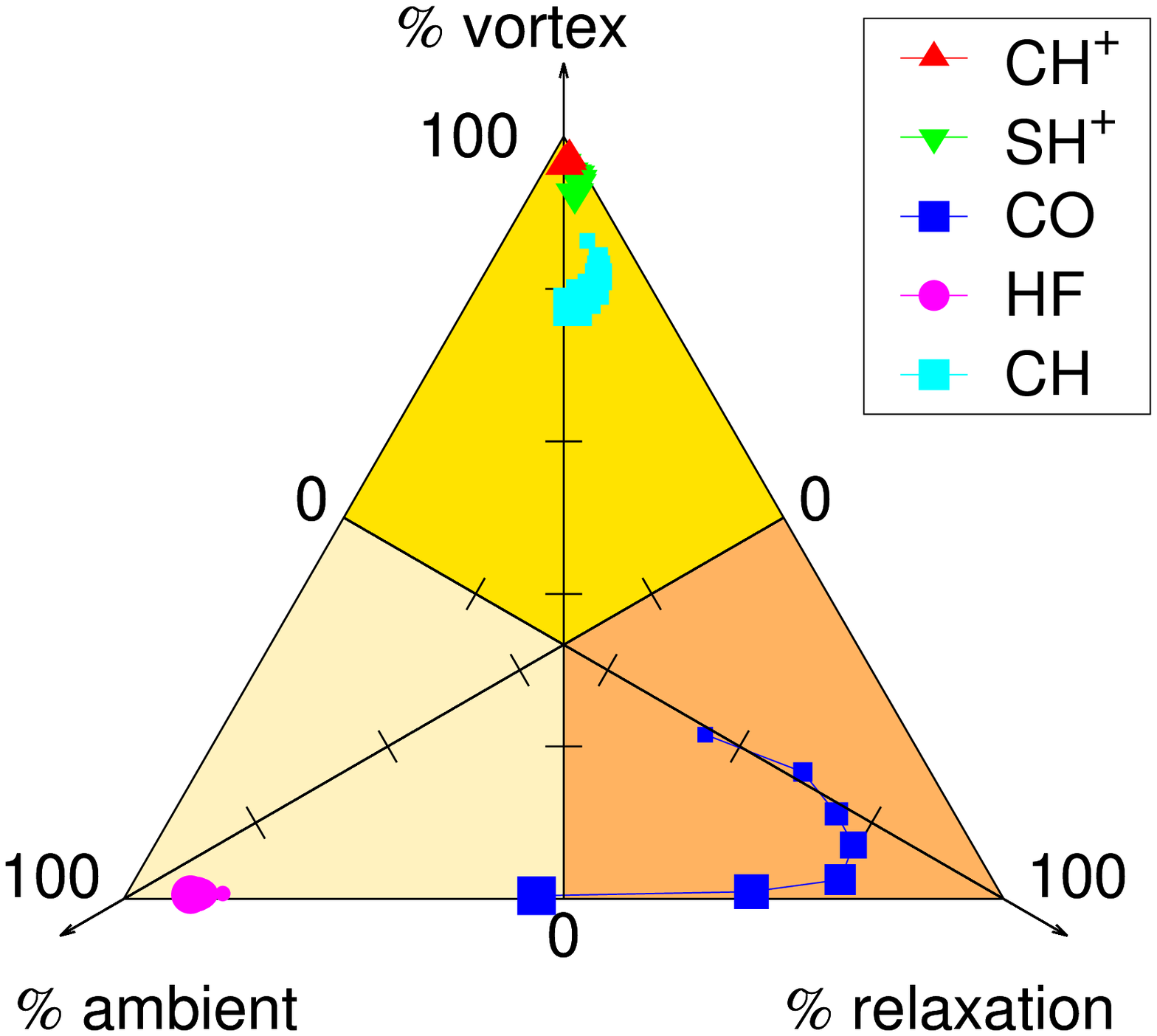}
\hspace{1.5cm} \rule{0.5mm}{6cm} \hspace{1cm}
\includegraphics[width=7.5cm,angle=-0]{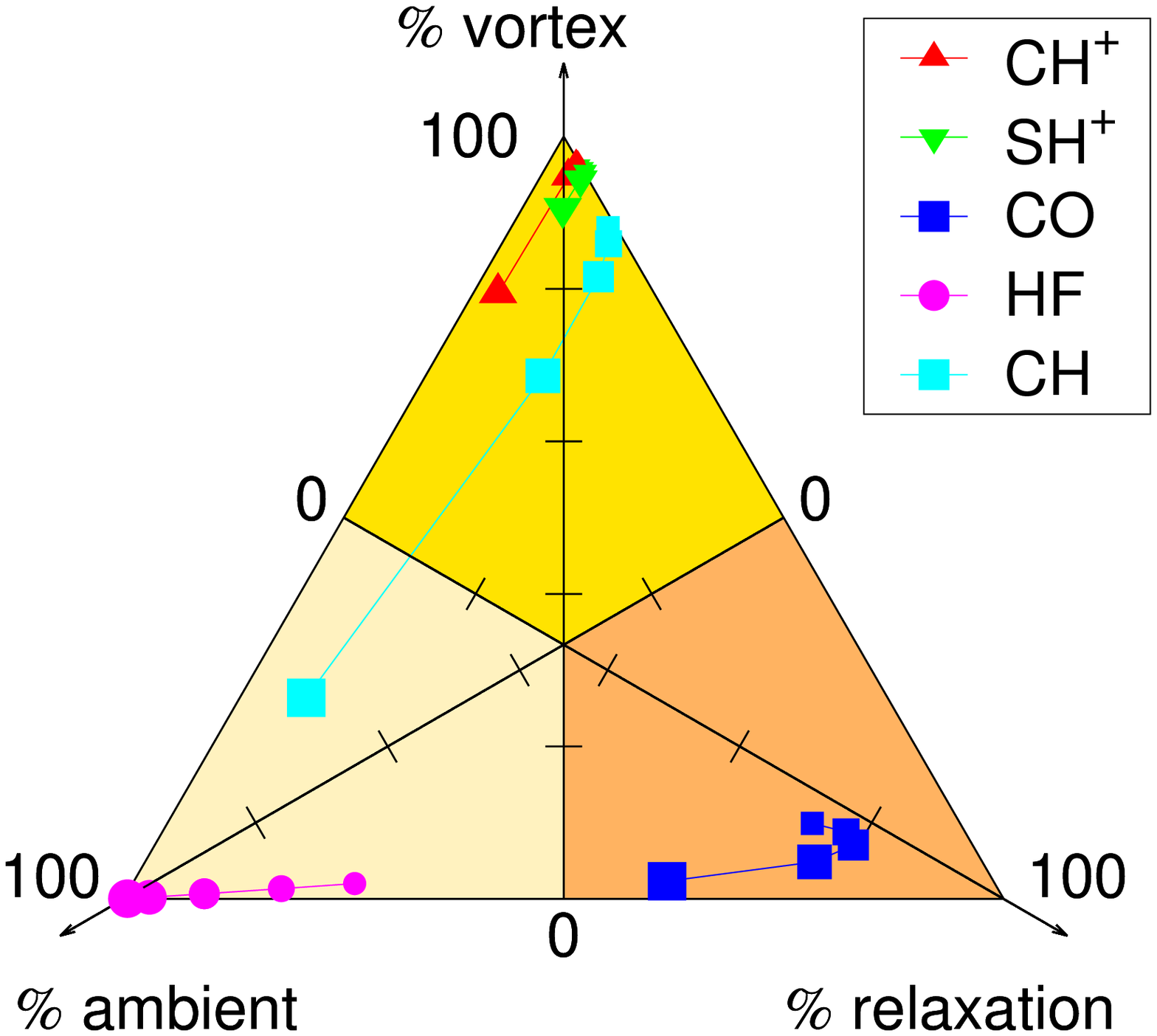}
\includegraphics[width=7.5cm,angle=-0]{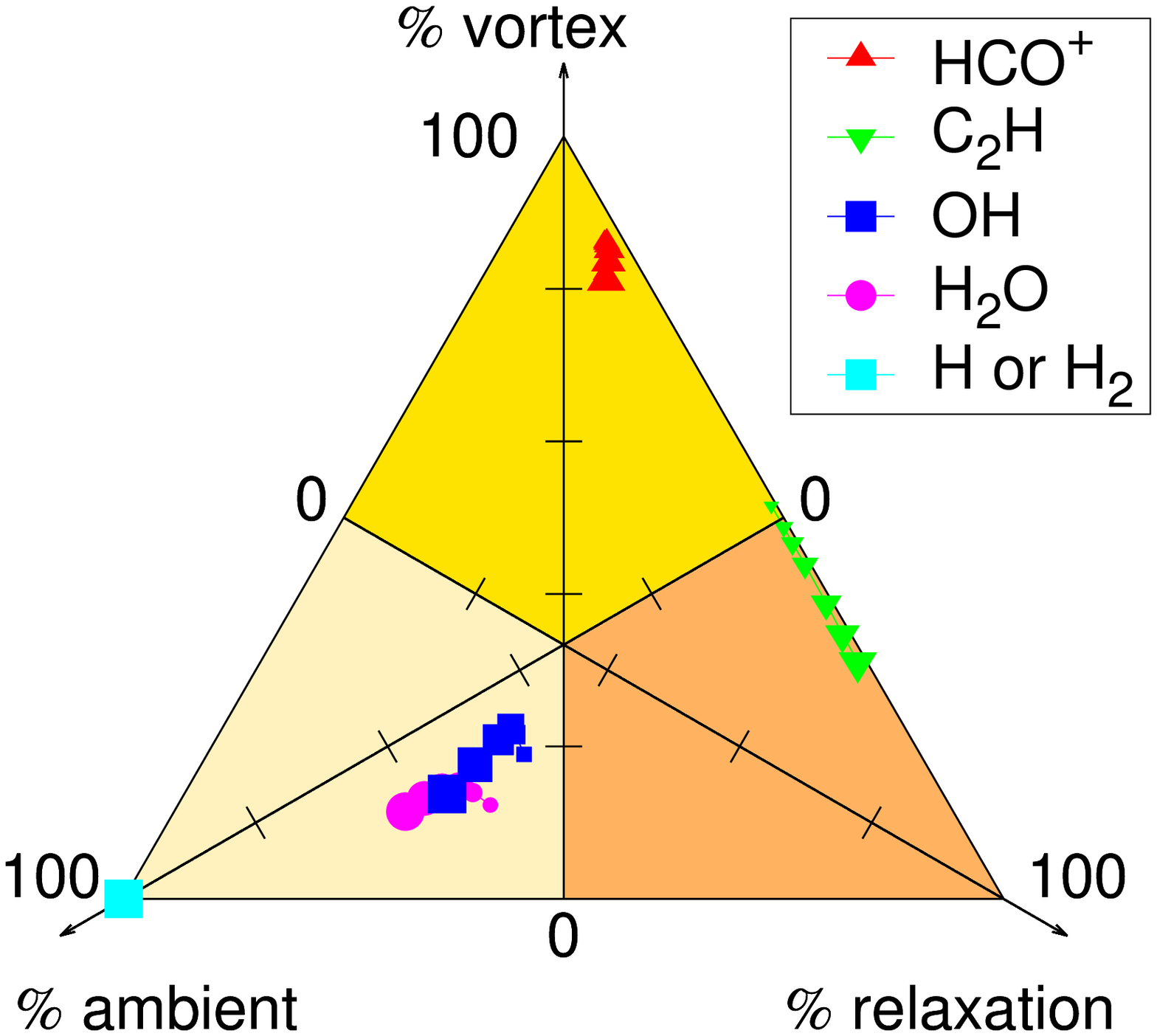}
\hspace{1.5cm} \rule{0.5mm}{7cm} \hspace{1cm}
\includegraphics[width=7.5cm,angle=-0]{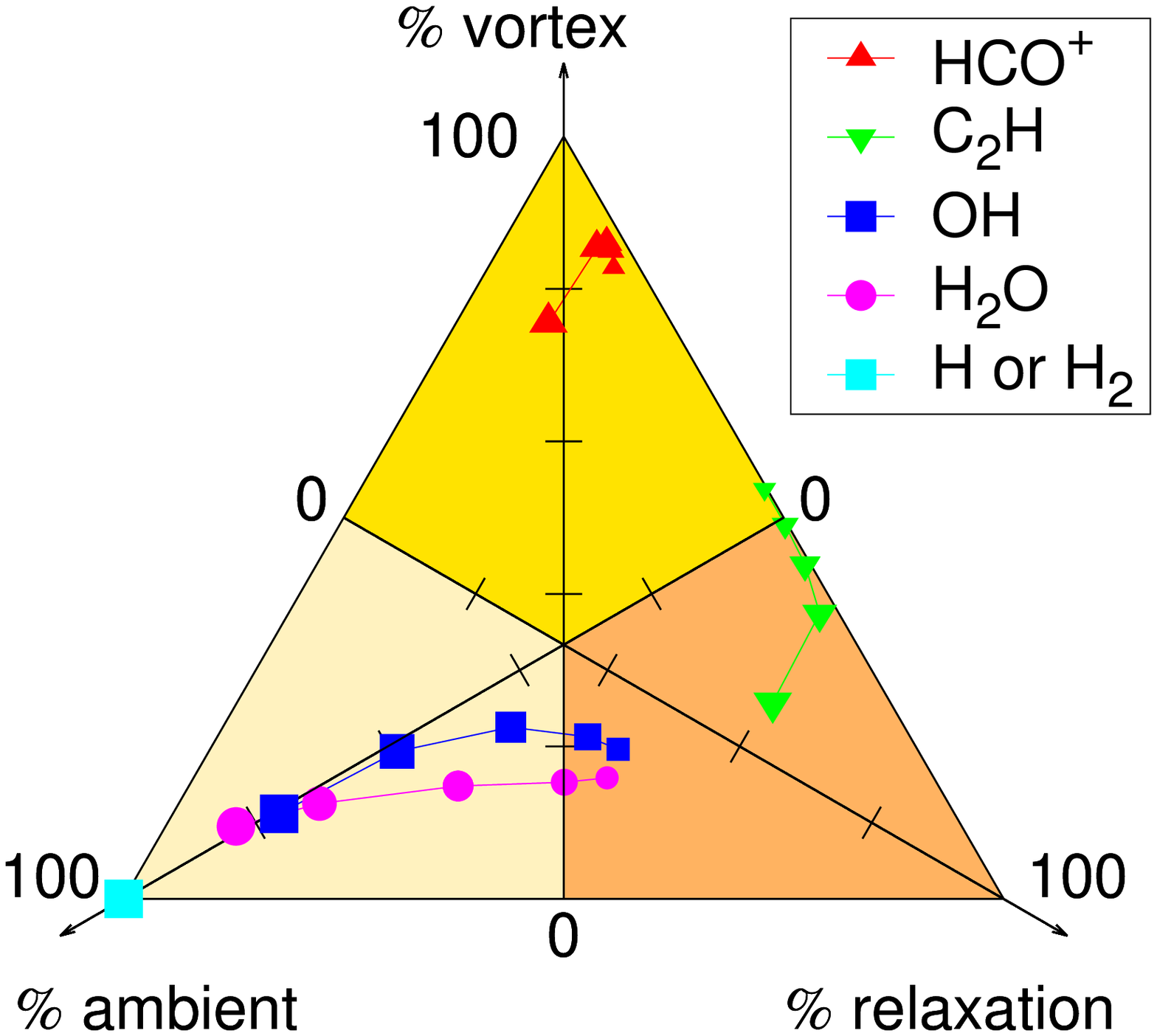}
\caption{Ternary phase diagram of the TDR model. The projections of each point on the
medians of the equilateral triangle show the contributions of each phase (ambient medium, 
active dissipative bursts, and relaxation stages) to the total column density of a given
species. In the left panels, the models are computed for several values of $A_V$: 
0.1, 0.2, 0.3, 0.4, 0.6, 0.8, and 1.0 mag. In the right panels, the models are computed 
for several values of $\dens$: 20, 30, 50, 100, and 300 \cc. All the other parameters are 
set to their standard values (see Table \ref{Tab-Par}). In all panels, the size of the 
symbols increases as the varying parameter increases. The beige, yellow, and orange zones 
indicate the regions where the production of the different species is dominated by the 
ambient medium, the active vortices, and the relaxation stages, respectively.}
\label{Fig-Triangles}
\end{center}
\end{figure*}

A helpful way to visualise the main results of the TDR model are the 
ternary phase diagrams shown in Fig. \ref{Fig-Triangles}. They display 
schematic representations of the contributions (see Eq. \ref{Eq-Coldens})
\begin{eqnarray}
f_{\rm amb}(X) & = & N_{\rm amb}({\rm X}) / N({\rm X}) \\[9pt]
f_V(X) & = & \mathscr{N}_V N_V({\rm X}) / N({\rm X}) \\[4pt]
f_R(X) & = & \mathscr{N}_V {\tau_R({\rm X}) \over \tau_V} {\overline N_R({\rm X})} / N({\rm X})
\end{eqnarray}
of the ambient medium, the active vortices, and the relaxation stages to the total column 
densities of ten molecules. These contributions are given by the projections of each point 
on the three triangle axis: for instance, the smallest symbol of CO (top left panel) reads 
$f_{\rm amb}({\rm CO}) = 22$ \%, $f_V({\rm CO}) = 21$ \%, and $f_R({\rm CO}) = 57$ \%.
These diagrams illustrate two major features of the model.

First, the phase contributions of the species selected here are distributed over the 
entire area of the triangles. It implies that each phase of the TDR model plays a role 
on the global chemical composition of the gas, even if it fills only a small fraction 
of the entire volume, and is therefore required to correctly model random diffuse 
lines of sight. In particular, Fig. \ref{Fig-Triangles} shows for the first time the 
critical impact of the relaxation times: the relaxation phase may account for 15 
\% of the total column density of CH, 50 \% of those of OH and \HdO, 70 \% of \CtH, and 
for more than 80 \% of the total column density of CO.

Second, except for CH, the production of which is dominated by the active phase at low 
density and by the ambient medium as the density increases, and for OH and \HdO, the 
productions of which may be dominated by the relaxation phase at low density, all the other 
species appear to be confined in specific regions of the triangles: H, \HH, and HF are 
almost entirely produced in the ambient medium; \SHp, \CHp, and \HCOp\ are predominantly 
formed in the active phase; while \CtH\ and CO mostly originate from the relaxation phase.
We therefore anticipate that the species produced in 
the active phase  can be used as signatures of the dynamics of the dissipation 
regions; those formed in the relaxation phase are expected to be sensitive to the relative 
timescales between the active dissipation and the relaxation; and those produced in the ambient 
medium are species unaffected by turbulent dissipation.

In the next sections, we select several pairs of species and compare both their column 
densities and column density ratios predicted by the TDR model to the observations of 
hundreds of diffuse lines of sight. The goal of this approach is to address the following 
questions: How do the model predictions compare with multiwavelength observations of atoms 
and molecules in the diffuse medium across the Galactic disk? Conversely, how can we use 
these observations to constrain the parameter space, hence infer the main properties of 
turbulent dissipation?

\section{\CHp\ and the turbulent dissipation rate} \label{Sect-CHp}

\begin{figure}[!hb]
\begin{center}
\includegraphics[width=8.5cm,angle=-0]{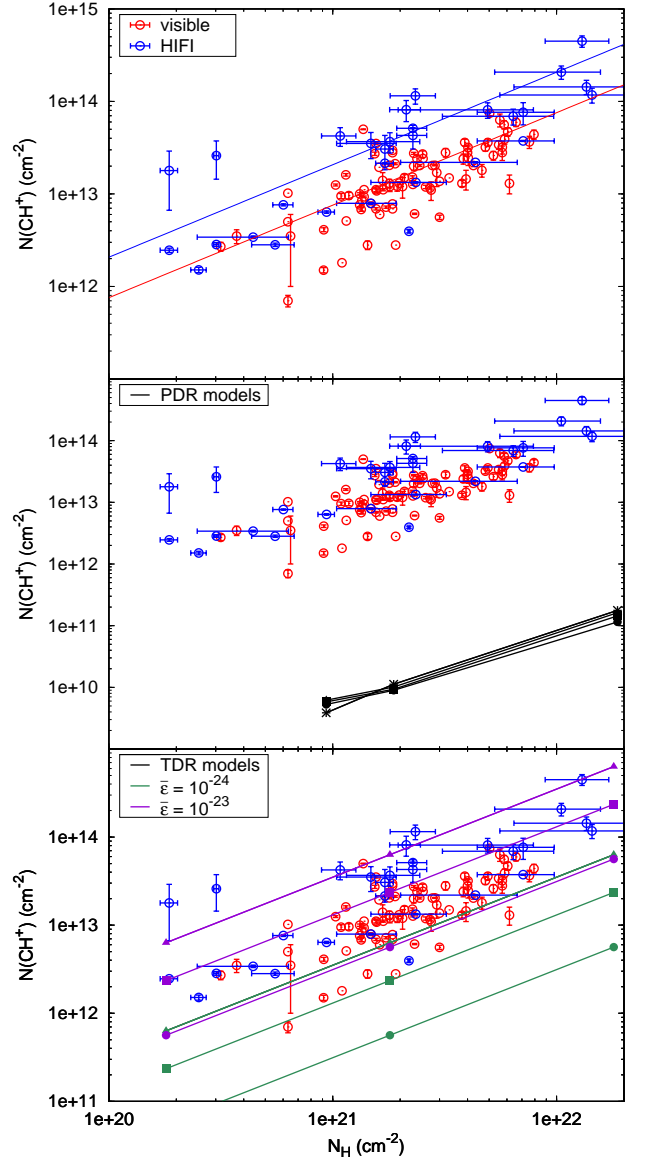}
\caption{Observations ({\it top panel}) compared to the predictions of PDR ({\it middle 
panel}) and TDR ({\it bottom panel}) models. Data (open circles) $-$ The \CHp\ and total 
hydrogen column densities inferred from visible absorption lines and extinction measurements 
are shown in red and are from \citet{Crane1995}, \citet{Gredel1997}, and \citet{Weselak2008}. 
The \CHp\ and total hydrogen column densities inferred from the far-infrared absorption 
lines of \CHp, HF, and CH and the 21 cm emission lines of HI are shown in blue and are 
from \citet{Falgarone2010}, \citet{Gerin2010}, \citet{Neufeld2010}, \citet{Sonnentrucker2010}, 
and \citet{Godard2012}. The red and blue lines correspond to the mean ratio $N(\CHp)/N_{\rm H}$ 
computed with the visible and submillimetre data, respectively. PDR and TDR model predictions 
(filled symbols) $-$ Computed for several densities: 10 (crosses), 30 (triangles), 50 (squares), 
and 100 \cc\ (circles). The TDR models are computed for $\epso = 10^{-24}$ (green) and 
$10^{-23}$ (magenta) \eccs, all the other parameters are set to their standard values.}
\label{Fig-CHp-NHtot}
\end{center}
\end{figure}


\subsection{Observational data} \label{Sec-obs-chp-H}

In the top panel of Fig. \ref{Fig-CHp-NHtot} we compare two datasets: 
the column densities of \CHp\ derived from its visible absorption lines (in red) detected
towards nearby OB stars \citep{Crane1995,Gredel1997,Weselak2008} and those deduced from
its far-infrared rotational transitions (in blue) recently observed in absorption towards 
the continuum of massive star forming regions with the \emph{Herschel}/HIFI instrument 
\citep{Falgarone2010,Godard2012}. The total hydrogen column densities associated with the
visible data are inferred from extinction measurements, while those associated with 
the far-infrared data are taken from \citet{Godard2012} who estimated the atomic hydrogen 
column density from VLA observations of its $\lambda$21 cm transition and the \HH\
column density from the far-infrared observations of two surrogate molecular species, CH
and HF \citep{Gerin2010,Neufeld2010,Sonnentrucker2010}. When both CH and HF are available, 
both estimations are used to compute uncertainties on the total column density, hence the 
large error bars on $N_{\rm H}$ for several far-infrared components.

Absorption spectroscopy against stars in the visible range is limited to the SN
 because of dust extinction, while that against star-forming regions in
the submillimetre domain does not have such a limitation (e.g. \citealt{Laor1993}). The red 
data of Fig. \ref{Fig-CHp-NHtot} are thus associated with gas in the SN 
while the blue data sample components located deeper in the inner Galactic disk (see 
Table 3 of \citealt{Godard2012}). These two datasets share similar characteristics: 
a large scatter of about one order of magnitude and the same linear 
trend with \NHt. However, the column densities of \CHp\ observed in the inner Galactic 
disk, with a mean ratio $\langle N({\rm CH}^+)/N_{\rm H} \rangle = 2.1 \times 10^{-8}$, 
are found to exceed by a factor of 3 on average those observed in the SN, 
where $\langle N({\rm CH}^+)/N_{\rm H} \rangle = 7.6 \times 10^{-9}$.

\subsection{Predictions of UV-dominated chemistry} \label{Sec-carb-chem}

The middle panel of Fig. \ref{Fig-CHp-NHtot} displays the column densities of \CHp\ predicted 
by the Meudon PDR code for densities of 30, 50, and 100 \cc, a total column density $\NHt= 
0.18$, $1.8$, and $18$ $\times 10^{21}$ \cq, and an incident radiation field scaling factor 
$\chi=1$. In the diffuse ISM, the destruction of \CHp\ occurs through hydrogenation (see Fig. 
\ref{Fig-Network-CHp} in Appendix \ref{Append-Networks}), an exothermic process with a very 
short timescale $\sim 1$ yr $f_{\HH} (50 \cc / \dens)$ \citep{McEwan1999}, where $f_{\HH}$ 
is the molecular fraction defined as $f_{\HH} = 2n(\HH) / \dens$. Conversely, in the absence 
of a suprathermal energy source to activate the endothermic reaction 
\begin{equation} \label{Reac-CHp-text}
\Cp +\HH \rightarrow \CHp + {\rm H} \quad (\Delta E / k \sim 4640 {\rm K}),
\end{equation} 
the formation of \CHp\ is initiated by the radiative association
\begin{equation}
\Cp +\HH \rightarrow \CHdp + \gamma,
\end{equation} 
a process with a very long timescale 
$\sim 2 \times 10^6$ yr $f_{\HH}^{-1} (50 \cc / \dens)$ \citep{Herbst1985}. With the lack of 
an efficient production pathway, PDR models thus systematically underestimate the observed 
column densities of \CHp\ by one to three orders of magnitude (Fig. \ref{Fig-CHp-NHtot}).

\subsection{Predictions of the TDR model} \label{Sect-Meas-epso}

In the bottom panel of Fig. \ref{Fig-CHp-NHtot}, we compare the observational data with the 
predictions of the TDR models. With a homogeneous reconstruction of the line of sight (see 
Sect. \ref{Sec-build-los}), all the column densities computed by the code are simply 
proportional to the total hydrogen column density $N_{\rm H}$. For a given $N_{\rm H}$, the 
dependence of $N(\CHp)$ on the other parameters is then the result of several effects. First, 
\CHp\ is predominantly produced in the active stages (see Fig. \ref{Fig-Triangles}); its 
column density is thus proportional to the number of active vortices, which is proportional 
to the average dissipation rate $\epso$, inversely proportional to the square of the density 
$\dens$ and independent of $E_{\tau_V}$ (see Paper I). Second, the rate-of-strain $a$ and the 
maximum orthoradial velocity $\utM$ control the heating by viscous friction and ion-neutral 
friction, respectively, hence set the effective temperature (Eq. \ref{Eq-Teff}) at which 
ion-neutral chemical reactions proceed. If $a$ or $\utM$ increase, the effective temperature 
in an active vortex increases and so does the rate of reaction \ref{Reac-CHp-text}, hence the
production of \CHp. When integrated over the line of sight, however, this effect is balanced 
by the number of vortices which decreases as the local dissipation rate increases (Paper I).
Third, since \CHp\ is formed through the hydrogenation of \Cp\ and the photodissociation of 
\CHtp\ (see Fig. \ref{Fig-Network-CHp} of Appendix \ref{Append-Networks}), its  
abundance, unlike most of the molecules, increases with the UV radiation field. With the 
analysis of all the models we find that the $N(\CHp)/N_{\rm H}$ column density ratio 
scales as
\begin{equation} \label{scaling-CHp}
N(\CHp) / N_{\rm H} \sim 1.5 \times 10^{-9}\, \overline{\varepsilon}_{24} \left( 
\frac{\dens}{50 \cc} \right)^{-2.2} \left( \frac{A_V}{0.4 {\rm mag}} \right)^{-0.3} \chi^{0.5}
\end{equation}
(with $\overline{\varepsilon}_{24} = 10^{24} \,\epso $ \eccs) 
for $\dens \lesssim 300$ \cc, and 
is almost independent of $\utM$ for $\utM \gtrsim 2.5$ \kms, i.e. as long as the gas rotation 
induces a sufficient ion-neutral drift to activate reaction \ref{Reac-CHp-text}.

The above relation shows a weak dependence of $N(\CHp)$ on both $A_V$ and $\chi$; $N(\CHp)$ 
varies by only a factor of 3 over the ranges covered by these two parameters which are 
inferred from the molecular fractions of the gas (see Sect. \ref{sec-data}) and from its 
positions in the Galactic disk (see Sect. \ref{Sect-Var-CHp}). Equation \ref{scaling-CHp} 
reveals stronger dependences of $N(\CHp)$ on both \epso\ and \dens, the latter reflecting 
the fast destruction of \CHp\ by collisions with hydrogen. This result somewhat hampers 
the interpretation of the observations: (1) there is a degeneracy between \epso\ and 
\dens\ from the sole observations of $N(\CHp)$ and $\NHt$ and (2) the steep dependence on 
\dens\ implies that the observations of \CHp\ preferentially sample the lowest densities 
of the diffuse medium.

The lowest values of $N(\CHp)/\NHt$ in Fig. \ref{Fig-CHp-NHtot} correspond to detection 
limits and thus cannot be used to infer a lower limit on the dissipation rate \epso\ or 
an upper limit on the gas density. In contrast, we can derive an upper limit of the turbulent 
dissipation rate from the largest observed \CHp\ abundances. The largest values of $N(\CHp)
/ \NHt$ observed in the local ISM (red points in Fig. \ref{Fig-CHp-NHtot}) and the inner 
Galactic disk (blue points in Fig. \ref{Fig-CHp-NHtot}) give $\overline{\varepsilon}_{24} 
\times (\dens / 50\cc )^{-2.2} \lesssim 20$ and 70, respectively. Assuming that these upper 
limits are associated with the lowest density components of the diffuse matter ($\dens \sim 
20$ \cc, \citealt{Fitzpatrick1997}, see also Sect. \ref{sec-data}), we find $\epso \lesssim 
10^{-23}$ \eccs\ for the components located in the inner Galactic disk, and about 3 times 
less in the SN. 

\subsection{Variation of $N(\CHp)/\NHt$ across the Galactic disk} \label{Sect-Var-CHp}

It is interesting to analyse the increase by a factor of three of the \CHp\ abundance between 
the SN and the inner Galaxy in the light of what is known about the Galactic dynamics and 
the star formation rate. Although the far-infrared lines of sight intercept the spiral arms 
of the Galaxy, we assume that, even in the inner Galaxy, the density of the gas rich in \CHp\
remains low because of the fast destruction rate of \CHp\ by collisions. The increase in the 
carbon elemental abundance towards the inner Galaxy accounts for an increase in the \CHp\ 
column density by only a factor of 1.2 at $R_g=6.5$ kpc because the trend is only of 0.037 
dex kpc$^{-1}$ \citep{Daflon2004}. The increase in the mean ISRF intensity towards the inner 
Galaxy has an influence \citep{Porter2005}. Using the SKY star distribution model \citep{
Wainscoat1992}, \citet{Moskalenko2006} estimate that the mean UV radiation field intensity 
increases by a factor of three from $R_g=8.5$ to 4 kpc. This sole increase, however, seems 
insufficient to explain the observed variation of the $N(\CHp)/N_{\rm H}$ ratio because most 
of the observed components beyond the SN lie in the range 5.5 $-$ 7.5 kpc \citep{Godard2012}.
What is left is an increase in the turbulent dissipation rate \epso\ that follows the 
turbulent injection rate. The latter scales with the supernovae explosion rate and the 
differential rotation of the Galaxy. These two processes are expected to increase \epso\ by 
a factor of 2.2 \citep{Lumsden2013,Urquhart2014} and 1.7 (assuming a flat Galactic rotation 
curve, \citealt{Fich1991}) from $R_g=8.5$ kpc to $R_g=6.5$ kpc. The increase in both the 
mean UV ISRF and the turbulent energy injection rate are therefore likely to be at the 
origin of the increase in the $N(\CHp)/\NHt$ ratio from the SN to the inner Galaxy.

\section{The \CHp\ to \SHp\ ratio and the ion-neutral velocity drift} \label{Sect-SHp}

\begin{figure}[!hb]
\begin{center}
\includegraphics[width=7.5cm,angle=-0]{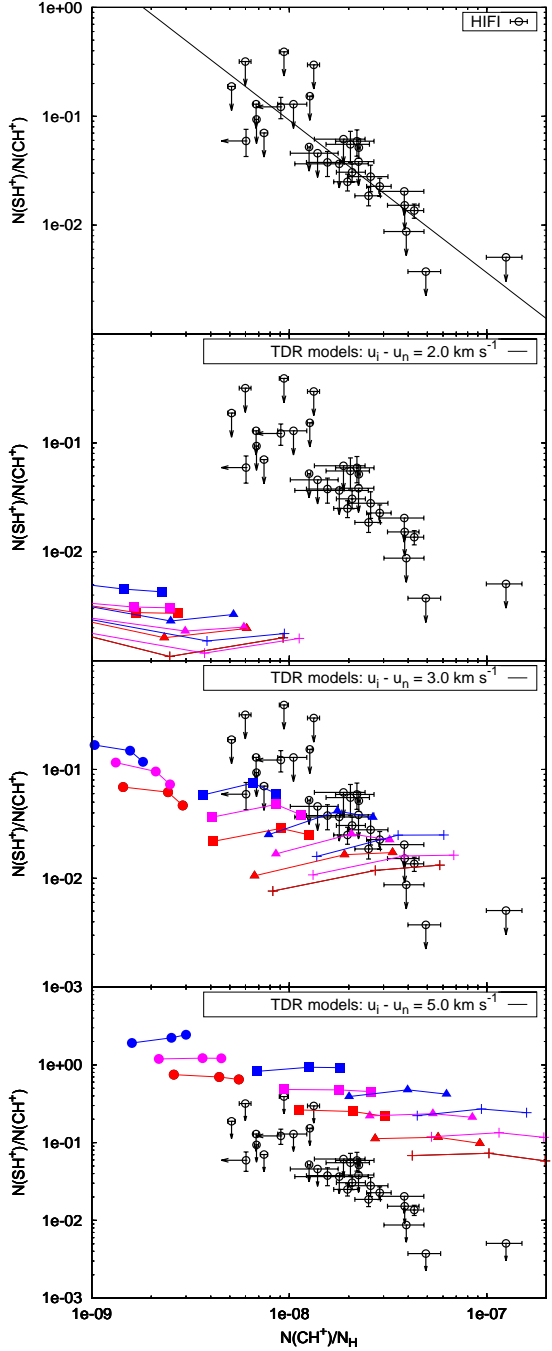}
\caption{Observations ({\it top panel}) compared to the predictions of TDR models computed
for three values of the maximum ion-neutral velocity drift: starting from the second panel, 
$u_i-u_n=2.0$, $3.0$, and $5.0$ \kms. Data (open circles) $-$ The \CHp\ 
and \SHp\ column densities and the least-squares power-law fit of their ratio (black line) 
are from \citet{Godard2012}. Model predictions (filled symbols) $-$ Computed for $A_V=
0.1$ (red), $0.2$ (magenta), and  $0.4$ (blue) mag and for several densities: 20 (crosses), 
30 (triangles), 50 (squares), and 100 \cc\ (circles). Along each curve, the rate-of-strain
$a$ varies between 10$^{-11}$ and $10^{-10}$ s$^{-1}$ from right to left.}
\label{Fig-SHp-CHp}
\end{center}
\end{figure}

We now select two species mainly formed in the active phases of the TDR model: CH$^+$ 
and SH$^+$ (see Fig. \ref{Fig-Triangles}).

\subsection{Observational data}

As the first step of the sulfur hydrogenation chain, the sulfanylium ion \SHp\ has long 
been sought without success in the local diffuse ISM through UV absorption spectroscopy 
of its electronic transitions \citep{Millar1988,Magnani1989,Magnani1991}. Only recently, 
thanks to the development of submillimetre and infrared telescopes (APEX, \emph{Herschel}), 
has interstellar \SHp\ finally been observed through its ground-state rotational transitions
\citep{Menten2011,Godard2012,Nagy2013}. Among these new data, those of \citet{Godard2012}
are particularly interesting because they show the first unambiguous detection of \SHp\
in diffuse interstellar components. As illustrated by the top panel of Fig. \ref{Fig-SHp-CHp}, 
the $\SHp/\CHp$ abundance ratio measured in the diffuse gas covers a broad range 
of values from less than 0.01 to more than 0.1 and appears to be proportional to $(N(\CHp)/
\NHt)^{-1.4}$. In this section we discuss the implications of this relation in the framework 
of UV-dominated and turbulence-dominated chemistries.

\subsection{Sulfur chemistry}

In many ways, the analysis of the sulfur hydrides provides unique constraints on chemical 
models applied to the interstellar medium. As quoted by \citet{Neufeld2012}, the sulfur 
differentiates itself from any other abundant element because its hydrides and hydride 
cations have small bonding energies. In consequence, the chemical reactions ${\rm X} + \HH 
\rightarrow {\rm XH} + {\rm H}$ are highly endothermic for ${\rm X} = {\rm S}$, ${\rm SH}$,
$\Sp$, $\SHp$, and $\SHdp$. The hydrogenation chain of sulfur therefore displays very 
distinctive chemical patterns when compared with that of carbon, both in a UV-driven and 
turbulence-driven regimes (see Fig. \ref{Fig-Network-CHp} in Appendix \ref{Append-Networks})

In a UV-driven chemistry, the sulfur hydrogenation is quenched by two reactions: 
\begin{equation} \label{Reac-SHdp-rad}
\Sp + \HH \rightarrow \SHdp + \gamma, \,\, {\rm and}
\end{equation}
\begin{equation} \label{Reac-SHtp-rad}
\SHp + \HH \rightarrow \SHtp + \gamma.
\end{equation}
With their very long characteristic timescales, these two reactions limit the production of 
SH and, by extension, the production of \SHp\ that originates from the photodissociation of 
\SHdp\ rather than the photoionisation of SH. In addition, and unlike \CHp\ and \CHdp\ 
which are destroyed by exothermic reactions with \HH, the destructions of \SHp\ and \SHdp\ 
mainly occur through
\begin{equation} \label{Reac-SHp-rec}
\SHp + e^- \rightarrow {\rm products}, \,\, {\rm and}
\end{equation}
\begin{equation} \label{Reac-SHdp-rec}
\SHdp + e^- \rightarrow {\rm products}.
\end{equation}
At equilibrium the $\SHp/\CHp$ abundance ratio writes
\begin{equation}
\frac{n(\SHp)}{n(\CHp)} \propto \frac{n(\Sp)}{n(\Cp)} \frac{n(\HH)^2}{n_e^2}
\end{equation}
and is independent from the extinction $A_V$ as long as \Cp\ and \Sp\ are the main ion 
carriers and the major reservoirs of carbon and sulfur. It results that the $\SHp/\CHp$ 
column density ratio integrated over PDR slabs is constant for a total extinction varying 
between 0.5 and 3 magnitudes and densities varying between 30 \cc\ and 500 \cc. This ratio 
is equal to 0.01 \citep{Godard2012}, i.e. between one and two orders of magnitude smaller 
than the largest observed value.

In a chemistry driven by turbulence, the release of energy due to dissipation is sufficient 
to activate the endothermic reactions producing the sulfur hydride cations such as
\begin{equation} \label{Reac-SHp-text}
\Sp +\HH \rightarrow \SHp + {\rm H} \quad (\Delta E / k \sim 9860 {\rm K}).
\end{equation}
The associated timescales decrease and become smaller than those of reactions \ref{Reac-SHdp-rad} 
and \ref{Reac-SHtp-rad}, yet still larger than those of reactions \ref{Reac-SHp-rec} and 
\ref{Reac-SHdp-rec}. Hydrogenation reactions therefore dominate the productions of \SHp, 
\SHdp, and \SHtp, while their destruction remain driven by the dissociative recombinations. 
From this simple scheme (see Fig. \ref{Fig-Network-CHp} in Appendix \ref{Append-Networks}), 
and using the endothermicities given in Eqs. \ref{Reac-CHp-text} and \ref{Reac-SHp-text}, 
the $\SHp/\CHp$ abundance ratio writes
\begin{equation}
\frac{n(\SHp)}{n(\CHp)} \propto \frac{n(\Sp)}{n(\Cp)} \frac{n(\HH)}{n_e} {\rm exp} \left( -\frac{5220 {\rm K}}{T_{\rm eff}} \right)
\end{equation}
at chemical equilibrium. It follows that $n(\SHp)/n(\CHp)$ strongly depends on the effective 
temperature $T_{\rm eff}$ of reactions \ref{Reac-CHp-text} and \ref{Reac-SHp-text}, thus on 
the velocity drift between $\Cp$ (or $\Sp$) and the molecular hydrogen (see Eq. \ref{Eq-Teff}).

\subsection{Influence of the TDR model parameters}

The three bottom panels of Fig. \ref{Fig-SHp-CHp} display a comparison between the
observational data and the predictions of the TDR models computed for $\utM=$ 2, 3, and 
5 \kms, a density ranging from 20 \cc\ to 100 \cc, a shielding varying between 0.1 mag 
and 0.4 mag, and a total dissipation rate $\epso = 3 \times 10^{-24}$ \eccs, i.e. within 
the range of values deduced in the diffuse ISM from the $N(\CHp)/\NHt$ ratio (cf. Sect. 
\ref{Sect-Meas-epso}).

As expected from the large difference between the \CHp\ and \SHp\ endothermicities, the 
$N(\SHp)/N(\CHp)$ column density ratio computed with the code is highly sensitive to the 
ion-neutral velocity drift set by the $\utM$ parameter. An increase in $\utM$ by a factor 
of two induces a rise of $N(\SHp)/N(\CHp)$ from one to three orders of magnitude. Detection 
limits on $N(\SHp)$ prevent us from deriving lower limits on \utM. In contrast, we infer 
$\utM \lesssim 3.5$ \kms\ regardless of $\dens$ and $A_V$. For $\utM = 3$ \kms, the code 
not only explains the most extreme values of the $N(\SHp)/N(\CHp)$ column density ratio, 
but also naturally reproduces its trend with $N(\CHp)/\NHt$. The dependence of $N(\CHp)$
and $N(\SHp)$ on the other parameters indicates that the matter traced by both \CHp\ and 
\SHp\ has a density $\dens \lesssim 50$ \cc\ in agreement with the results of the previous 
section.

\section{The CO to \HCOp\ ratio and the dissipation timescale} \label{Sect-CO}

We now focus on \HCOp and CO, i.e. on two species which, according to the ternary phase 
diagrams (Fig. \ref{Fig-Triangles}), originate from the active and relaxation stages, 
respectively. Because these two species are closely related by their chemistry and belong 
to two different phases of the TDR model, they offer a unique opportunity to obtain 
information on the relative value of the relaxation and active dissipation timescales.

\subsection{Observational data}

\begin{figure*}[!ht]
\begin{center}
\includegraphics[width=16.0cm,angle=-0]{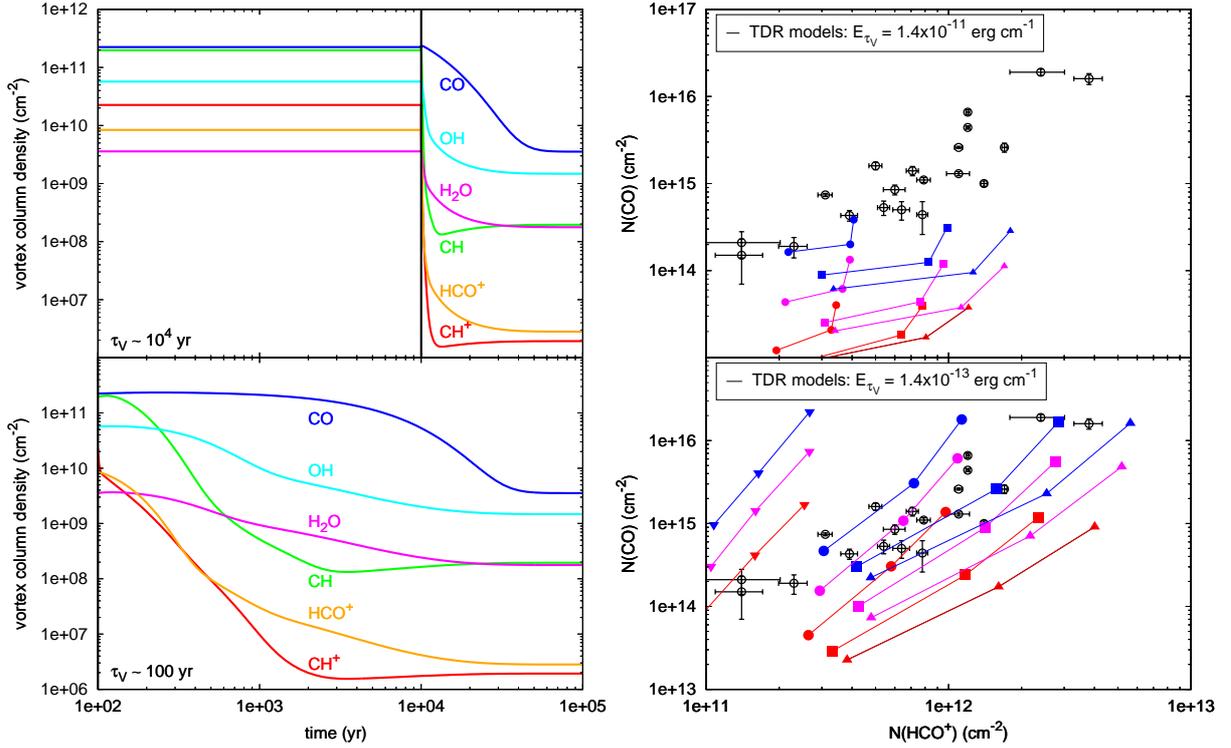}
\caption{{\it Left panels}: column densities of selected species integrated across a single vortex 
as functions of time during the isochoric relaxation phase. The models are computed for two 
values of the vortex dissipation integral $E_{\tau_V}=1.4 \times 10^{-11}$ \ecm\ ({\it top 
panels}) and $E_{\tau_V}=1.4 \times 10^{-13}$ \ecm\ ({\it bottom panels}). The corresponding 
vortex lifetimes are indicated with vertical straight lines. All the other parameters are 
set to their standard values. {\it Right panels}: comparison of observations with the 
predictions of TDR models. The observed CO and \HCOp\ column densities (open circles) are 
from \citet{Lucas1996,Liszt1998,Liszt2004a}. The predictions of the TDR models (filled 
symbols) are shown for $A_V= 0.2$ (red), $0.4$ (magenta), and $0.8$ (blue) mag, 
$\dens = 30$ (triangles), $\dens = 50$ (squares), $\dens = 100$ (circles), and 
$\dens = 300$ \cc\ (upside-down triangles), and a total column density $\NHt = 1.8 
\times 10^{21} \cq$; along each curve, the rate-of-strain $a$ varies between 
10$^{-11}$ (right) and $10^{-10}$ s$^{-1}$ (left).}
\label{Fig-CO-HCOp}
\end{center}
\end{figure*}


Two different data sets are used: (1) the \HCOp(1-0) and CO(1-0) lines observed in absorption 
on the continuum of quasars \citep{Lucas1996,Liszt1998,Liszt2004a} and (2) the CO and \HH\ 
lines observed in the visible and UV ranges in absorption towards nearby OB stars (e.g. 
\citealt{Burgh2007,Sheffer2008}).

In the first set of data (Fig. \ref{Fig-CO-HCOp}), the selected lines of sight sample all 
Galactic latitudes (\citealt{Liszt2010}, Appendix). The observed column densities of CO 
and \HCOp\ show a non-linear correlation $N({\rm CO}) \propto N(\HCOp)^{1.4}$ with an 
averaged abundance ratio varying between $5 \times 10^2$ and $8 \times 10^{3}$. The total 
column densities associated with each data point are similar and close to one magnitude of 
matter (Paper I). 

In the second set of data, displayed in the top panel of Fig. \ref{Fig-CO-NHtot}, the 
sightlines sample diffuse components located in the SN, i.e. within $\sim 1$ kpc from 
the sun. No correlation is observed between $N({\rm CO})$ and $N(\HH)$. Unlike in the 
previous data set, the (unknown) total hydrogen column density associated with the CO 
data likely spans a large range, not only because the \HH\ column densities vary by two 
orders of magnitude, but also because the \HH\ fraction is expected to fluctuate by a 
large factor (Sect. \ref{sec-data}).

\subsection{Oxygen chemistry} \label{Sec-oxy-chem}

In diffuse interstellar gas, CO is expected to be destroyed by photodissociation and
formed by the dissociative recombination of \HCOp, regardless of 
turbulent dissipation (see Fig. \ref{Fig-Network-CO} of Appendix \ref{Append-Networks}). 
At chemical equilibrium, the CO/\HCOp\ abundance ratio writes
\begin{equation} \label{Eq-CO-HCOp}
\frac{n({\rm CO})}{n(\HCOp)} = \frac{n_e k_e}{k_\gamma}, 
\end{equation}
where $k_e = 2.46 \times 10^{-7} (T/300 \,\,{\rm K})^{-0.69}$ cm$^{3}$ s$^{-1}$ 
\citep{Florescu-Mitchell2006}. Adopting the photodissociation rate $k_\gamma$ (in s$^{-1}$)
of CO given in Fig. \ref{Fig-Photo-CO} (Appendix \ref{Append-H2-CO}), $n({\rm CO})/n(\HCOp)$ 
is found to vary between $10^2$ and $10^5$ for typical diffuse gas conditions ($30 \lesssim 
\dens \lesssim 300$ \cc, $0.01 \lesssim A_V \lesssim 1$ mag), i.e. to entirely cover the 
observed range (Fig. \ref{Fig-CO-HCOp}). Consequently, any model computing the chemical 
equilibrium of the gas should be able to explain the observed abundance of CO, provided 
that it produces enough HCO$^+$.

In a UV-dominated chemical scheme (see Fig. \ref{Fig-Network-CO} of Appendix 
\ref{Append-Networks}), the formation of \HCOp\ proceeds though the oxygen 
hydrogenation chain followed by the two reactions: ${\rm OH} + 
\Cp \rightarrow \COp +{\rm H}$ and $\HdO + \Cp \rightarrow \HCOp +{\rm H}$. 
The bottleneck process of this pathway is the conversion of O into \Op\ via 
the charge exchange with ionised hydrogen. The abundance of \HCOp\ therefore 
depends on that of \Hp, the production of which is driven by the cosmic ray ionisation 
of H and \HH. To investigate this effect, we have run the Meudon PDR code along 
the grid presented in Appendix \ref{Append-H2-CO} assuming two values of the 
cosmic ray ionisation rate of molecular hydrogen : $\zetaHH = 3$ and $30 \times 
10^{-17}$ s$^{-1}$. In all cases, and despite the strong dependence of $n(\HCOp)$ 
on $\zetaHH$, we find that the PDR models systematically underestimate the column 
densities of \HCOp\ by more than one order of magnitude for $\zetaHH = 3 \times 
10^{-17}$ s$^{-1}$ and by more than a factor of 5 for $\zetaHH = 3 \times 10^{-16}$ 
s$^{-1}$. These results are in line with those of Paper I and \citet{Levrier2012} 
who found similar discrepancies by comparing the predictions of the Meudon PDR code 
with the observed column densities of CO.

In an active vortex (bottom panel of Fig. \ref{Fig-Network-CO} of Appendix \ref{Append-Networks}) 
it is not the hydrogenation chain of O that matters, but that of C. Indeed it enhances the 
production of the very reactive cation, \CHtp, which swiftly reacts with O to form \COp\ 
and \HCOp. With the opening of this additional formation pathway, the abundance of \HCOp\ 
rises by several orders of magnitude and that of CO by a factor of up to 100 (see Fig. 
3 of Paper I).  If these behaviors appear to be in contradiction with Eq. \ref{Eq-CO-HCOp}, 
it is because in dissipative regions the chemistry is not at equilibrium and the electronic 
fraction is substantially smaller than that in the ambient gas. After integration over the 
entire line of sight, we showed in Paper I  that the sole contributions of the active phases 
could explain the observed column densities of \HCOp\ for small values of the rate-of-strain 
$10^{-11} \lesssim a \lesssim 10^{-10}$ s$^{-1}$, but fail to reproduce those of CO by more 
than one order of magnitude. Because \HCOp\ and CO only differ by the timescales associated 
with their respective destruction processes, we suggested in Paper I that CO may in fact trace 
the relaxation period of turbulent dissipation (Fig. \ref{Fig-Triangles}).

\subsection{Influence of the TDR model parameters} \label{Sec-oxy-TDR}

To illustrate the  influence of the relaxation period for CO, we display the column densities 
of several species computed across the diameter of a single vortex as functions of time in 
the left panels of Fig. \ref{Fig-CO-HCOp}. All parameters are set to their standard values 
except for $E_{\tau_V}$ which is set to $1.4 \times 10^{-11}$ \ecm\ (top panels) and $1.4 
\times 10^{-13}$ \ecm\ (bottom panels). The corresponding values of the vortex lifetime (see 
Eq. \ref{Eq-Etau}) are indicated by straight vertical lines. Because the vortex is stationary, 
the column densities it produces are constant over its entire lifetime ($t \leqslant \tau_V$). 
Once the vortex has destabilised ($t > \tau_V$), the column density of each species decreases 
towards that produced by the equivalent amount of matter of ambient gas with a characteristic 
timescale that depends on the species.

Since the computed column density of a given species takes its time-dependent decrease 
into account during the relaxation phase (Eq. \ref{Eq-Coldens}), Fig. \ref{Fig-CO-HCOp} 
illustrates the importance of the relative values of the durations of the active and 
relaxation phases and their dependence on the vortex dissipation integral $E_{\tau_V}$.
For large values, $E_{\tau_V} = 1.4 \times 10^{-11}$ \ecm, i.e. large values of $\tau_V$, 
the relaxation phase has no impact on the time-averaged column density of any species. For 
small values, $E_{\tau_V} = 1.4 \times 10^{-13}$ \ecm, i.e. small values of  $\tau_V$, the 
relaxation phase dominates the time-averaged column densities of CO, OH, and \HdO, and may 
start to have an impact on those of \HCOp\ and CH. 

The resulting total column densities of CO and \HCOp\ are displayed in the right panels of 
Fig. \ref{Fig-CO-HCOp}. We find that the TDR model can account for the observed ranges of
column densities of both CO and \HCOp\ if $E_{\tau_V}$ is sufficiently small because 
$E_{\tau_V}$ affects CO and \HCOp in different ways: $N({\rm CO})$ increases by a factor 
of 60 as $E_{\tau_V}$ decreases by a factor of 100, while $N(\HCOp)$ only increases by a 
factor of 3. Further reducing $E_{\tau_V}$ increases both $N({\rm CO})$ and $N(\HCOp)$. 
These effects are due to the large difference between the relaxation times of \HCOp\ and 
CO that affect differently the total column densities, depending on the value of $\tau_V$ 
(Eq. \ref{Eq-Coldens}). The best range of vortex integral dissipations is  $10^{-13} 
\lesssim E_{\tau_V} \lesssim 10^{-12}$ \ecm, which corresponds to a vortex lifetime 
ranging between 100 yr and 1000 yr (for the standard vortex).

Interestingly, \HCOp\ and CO depend differently on the other parameters. Both $N({\rm CO})$ 
and $N(\HCOp)$ decrease as $a$ increases, \HCOp\ is most sensitive to $\dens$, but is almost 
independent of $A_V$, while the reverse is true for CO. Adopting the best range for $E_{\tau_V}$, 
we find that the TDR model reproduces the observations for all the values $10^{-11} \leqslant 
a \leqslant 10^{-10}$ s$^{-1}$, in agreement with our previous work (Paper I). It is predicted 
that the detections of \HCOp\ and CO trace a medium with a density comprised between 30 \cc\
and 300 \cc\ and a shielding $A_V \gtrsim 0.2$ mag, i.e. a gas phase denser and more shielded 
from the UV photons than that inferred from the observations of \CHp\ and \SHp.

\subsection{Highly fragmented versus PDR-like medium }

\begin{figure}[!ht]
\begin{center}
\includegraphics[width=8.5cm,angle=-0]{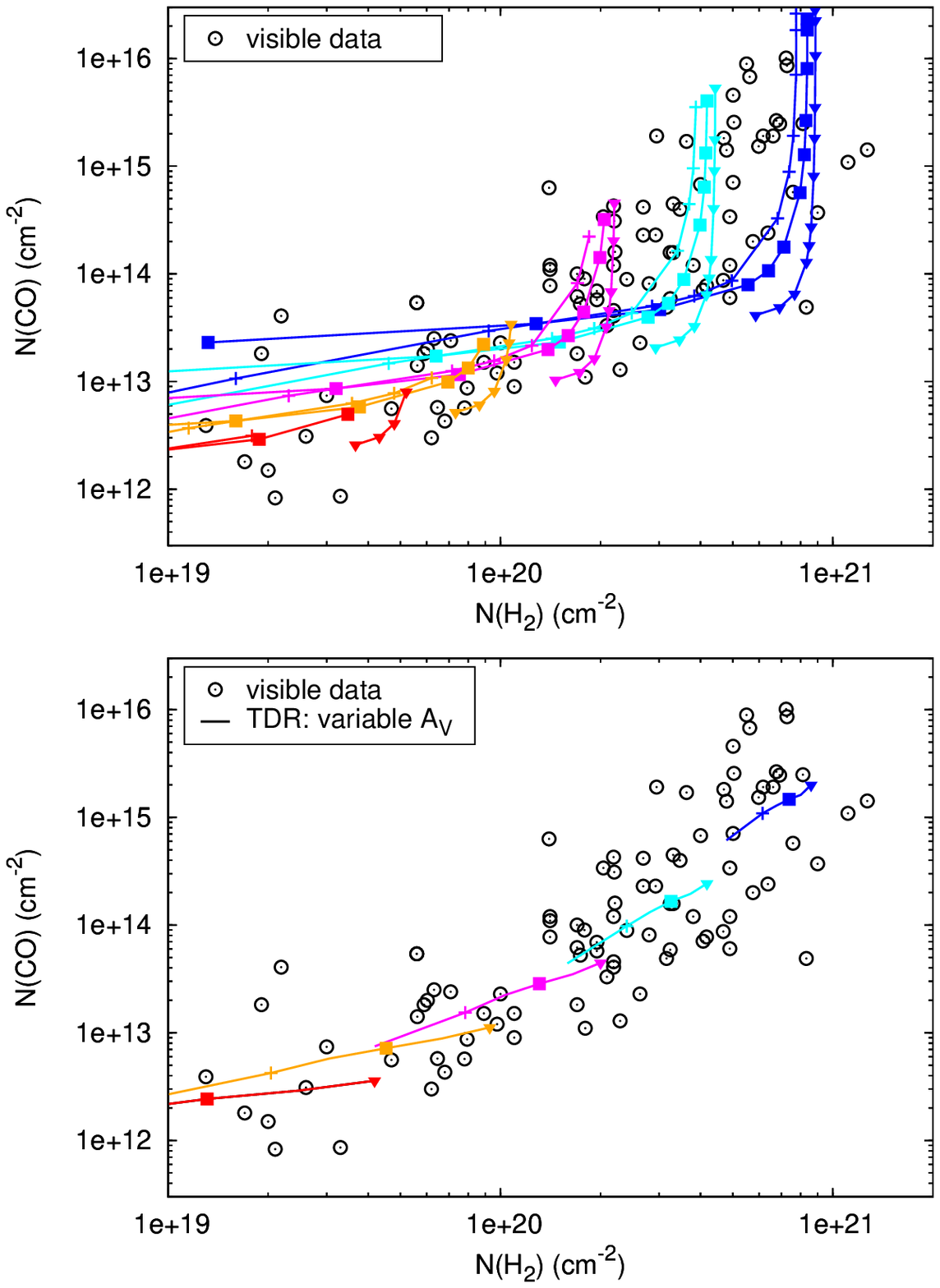}
\caption{Observations of the column densities of CO and \HH\ (open circles, \citealt{Sheffer2008})
compared to the predictions of the TDR model. The model (filled symbols) are computed for several 
densities, $\dens = 20$ (crosses), 50 (squares), and 300 \cc\ (upside-down triangles) and total 
column densities, $\NHt = 1.1$ (red), $2.2$ (orange), $4.5$ (magenta), $9.0$ (cyan), and $18$ 
(blue) $\times 10^{20}$ \cq. {\it Top panel:} the lines of sight are reconstructed with a
homogeneous shielding. The value adopted for $A_V$ varies along each curve from left to right 
from 0.005 mag to $A_{V,{\rm max}}$, where $A_{V,{\rm max}}$ is the shielding equivalent to a 
column density of matter of $\NHt/2$. {\it Bottom panel:} the lines of sight are reconstructed
with a PDR-like shielding function of $\NHt$ (Sect. \ref{Sec-build-los}), hence $A_V$ is no 
longer a free parameter.}
\label{Fig-CO-NHtot}
\end{center}
\end{figure}

The predictions of the TDR model are now compared with the column densities of CO and \HH\ 
observed in the SN, i.e. two species sensitive to the porosity of the gas to UV radiation 
(Fig. \ref{Fig-CO-NHtot}). The model parameters are set to their standard values. We adopt 
the two prescriptions for the extinction of the UV field discussed in Sect. \ref{Sec-build-los}: 
a homogeneous shielding along the line of sight (top panel) and a PDR-like shielding increasing 
with $\NHt$ (bottom panel). In both cases, we explore the impact of very small UV shielding 
(as small as 0.005 mag) on the column densities of CO predicted by the model.

The results shown in Fig. \ref{Fig-CO-NHtot} confirm the results reported in the previous
section and extend towards low column densities. They are in good  agreement with the 
observations even for $N(\HH)$ as small as $10^{19}$ \cq. They also illustrate the influence 
of the fragmentation of the gas on $N({\rm CO})$ and $N(\HH)$.

The predictions obtained with a highly fragmented medium (top panel) reproduce both the 
general increase in $N({\rm CO})$ with $N(\HH)$ and the large scatter of the data (about a 
factor of 10), whatever the other parameters. In this case, large column densities of CO 
at small $N(\HH)$ are explained by the accumulation of many fragments, each weakly shielded 
from the UV  field. We stress that $A_V$ as low as 0.005 mag combined with large \NHt\ might 
reproduce the largest CO column densities at low $N(\HH)$. A PDR-like reconstruction of 
the line of sight (bottom panel) reproduces the increase of $N({\rm CO})$ on $N(\HH)$ and 
especially the flattening at small $N(\HH)$. The $N({\rm CO}) / N(\HH)$ column density 
ratio is almost independent of the gas density. In this case, the observed scatter can 
only be explained by invoking variations of the other parameters over the acceptable 
ranges inferred in the previous sections.

Turbulent dissipation is therefore a crucial process to explain the CO abundances observed 
in the diffuse gas. The fragmentation of the gas and its permeability to UV photons has an 
important impact on the column densities of CO and molecular hydrogen. The two prescriptions 
discussed here (top and bottom panels of Fig. \ref{Fig-CO-NHtot}), however, are plausible 
scenarios, although the arguments developed in Sect. 2.3 on the broad width of the H/\HH\ 
transition region tend to favour a high level of fragmentation of the diffuse matter.

\section{[CII] line emission versus far-infrared dust emission} \label{Sect-Cp}

\subsection{Observational data}

\begin{figure}[!ht]
\begin{center}
\includegraphics[width=8.5cm,angle=-0]{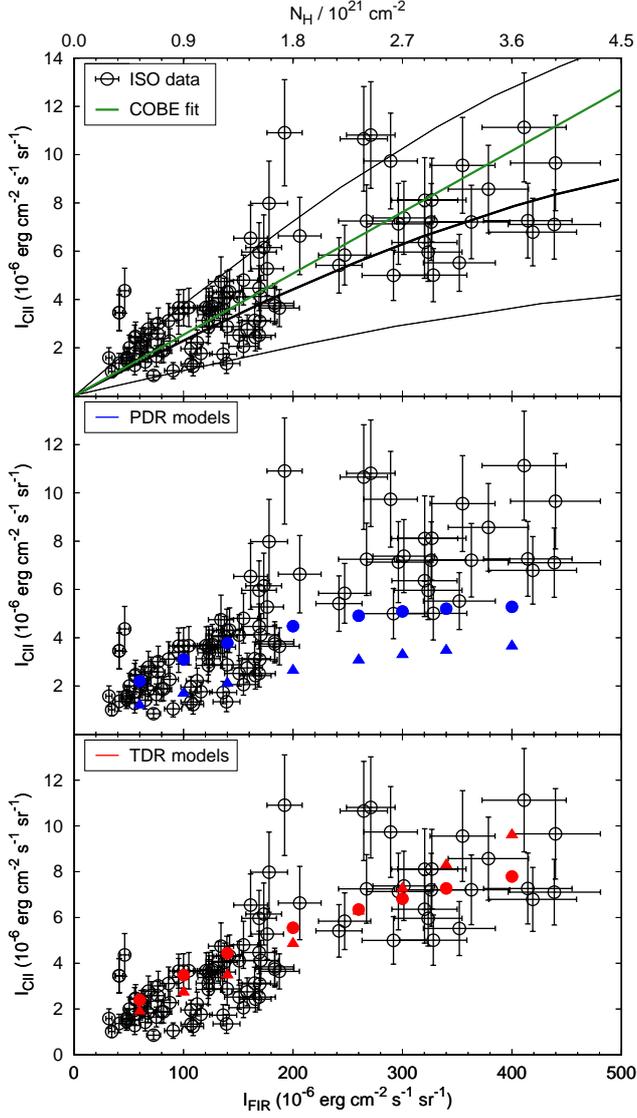}
\caption{Intensities of the [CII] and of the far-infrared dust 
emission observed by \citet{Ingalls2002} at high Galactic latitudes (black empty circles)
and comparison with the predictions of PDR ({\it middle panel}) and TDR ({\it bottom panel}) 
models. The black curves ({\it top panel}) are the PDR predictions of \citet{Ingalls2002} 
assuming a photoelectric heating efficiency of 7 \%, 4.3 \%, and 2 \% from top to bottom. 
PDR and TDR models (blue and red filled symbols) are computed for $a= 10^{-11}$ s$^{-1}$
and $\dens= 30$ (triangles) and 100 \cc\ (circles), all other parameters are set to their 
standard values.}
\label{Fig-Cp-FIR}
\end{center}
\end{figure}


In this section we focus on the 158 $\mu$m [CII] line observations performed by \citet{
Ingalls2002} towards 101 transluscent lines of sight at high Galactic latitudes. Their 
observational results are displayed in Fig. \ref{Fig-Cp-FIR} as a function of the far-infrared 
dust brightness. They show that (1) the [CII] line and the interstellar dust emissions are 
linearly correlated $I_{\rm [CII]} = (2.5 \pm 0.9) \times 10^{-2} I_{\rm FIR}$, in agreement 
with the results of the COBE Galactic survey \citep{Wright1991,Bennett1994}; (2) this relation 
breaks for $I_{\rm FIR} \gtrsim 2.5 \times 10^{-4}$ \ecqs\ indicating a slower increase in 
[CII] emission at high values of far-infrared brightness (i.e. at high column densities of 
matter); and (3) the observed scatter is real.

\citet{Ingalls2002} interpreted these data in terms of a variable efficiency of the dust 
photoelectric heating in diffuse gas. Assuming that the cooling by [CII] entirely balances 
the heating by the UV photons, as expected in PDR environments, they deduced the heating 
rate in order to match the [CII] emission. Their best-fit scenario (thick black curve on 
the top panel of Fig. \ref{Fig-Cp-FIR}) is obtained for an efficiency of 4.3 \% and a 
radiative energy flux of 1.6 times the standard value in the SN \citep{Habing1968}. While 
plausible, this interpretation raises an issue: the best-fit value is larger than the 3 \% 
obtained by \citet{Bakes1994} with the COBE Galactic survey and the range of efficiencies 
needed to reproduce the scatter exceeds the 5 \% limit they prescribed.

\subsection{Predictions of PDR models}

The middle panel of Fig. \ref{Fig-Cp-FIR} displays a comparison between the observations
and the predictions of the Meudon PDR code computed for two densities $\dens=30$ and 
$\dens=100$ \cc, a radiation scaling factor $\chi=1$, and a total amount of matter varying 
between 0.3 and 2 magnitudes. We adopt the conservative value of 3 \% for the photoelectric 
heating efficiency. As found by \citet{Ingalls2002}, a PDR-type medium naturally explains 
the slower increase in \Cp\ emission at large $I_{\rm FIR}$ because at large visible extinction 
($A_V \gtrsim 1$ mag) the gas temperature is too low to excite the $^2P_{3/2} \rightarrow 
^2P_{1/2}$ transition of \Cp\ and because \Cp\ undergoes the \Cp/C/CO transition. Once the 
transition is completed, the integrated emission of \Cp\ across the slab tends towards a 
constant value regardless the total amount of matter along the line of sight. Yet, the 
comparison with the observations shows that PDR models computed with a photoelectric effect 
efficiency of 3 \% underestimate by a factor 2 to 3 the observed \Cp\ emission for 
$I_{\rm FIR} \gtrsim 2 \times 10^{-4}$ \ecqs.


\subsection{Predictions of TDR models}

\begin{figure}[!ht]
\begin{center}
\includegraphics[width=8.5cm,angle=-0]{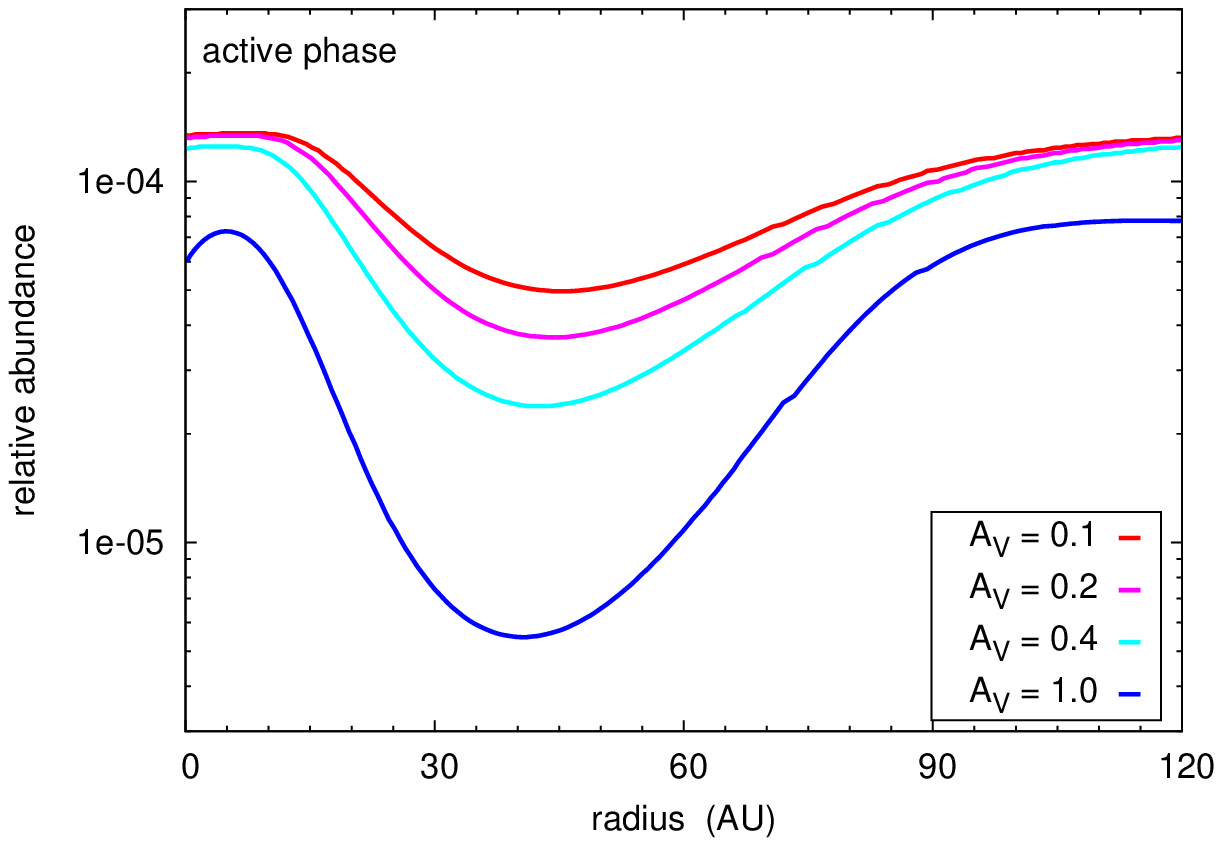}
\includegraphics[width=8.5cm,angle=-0]{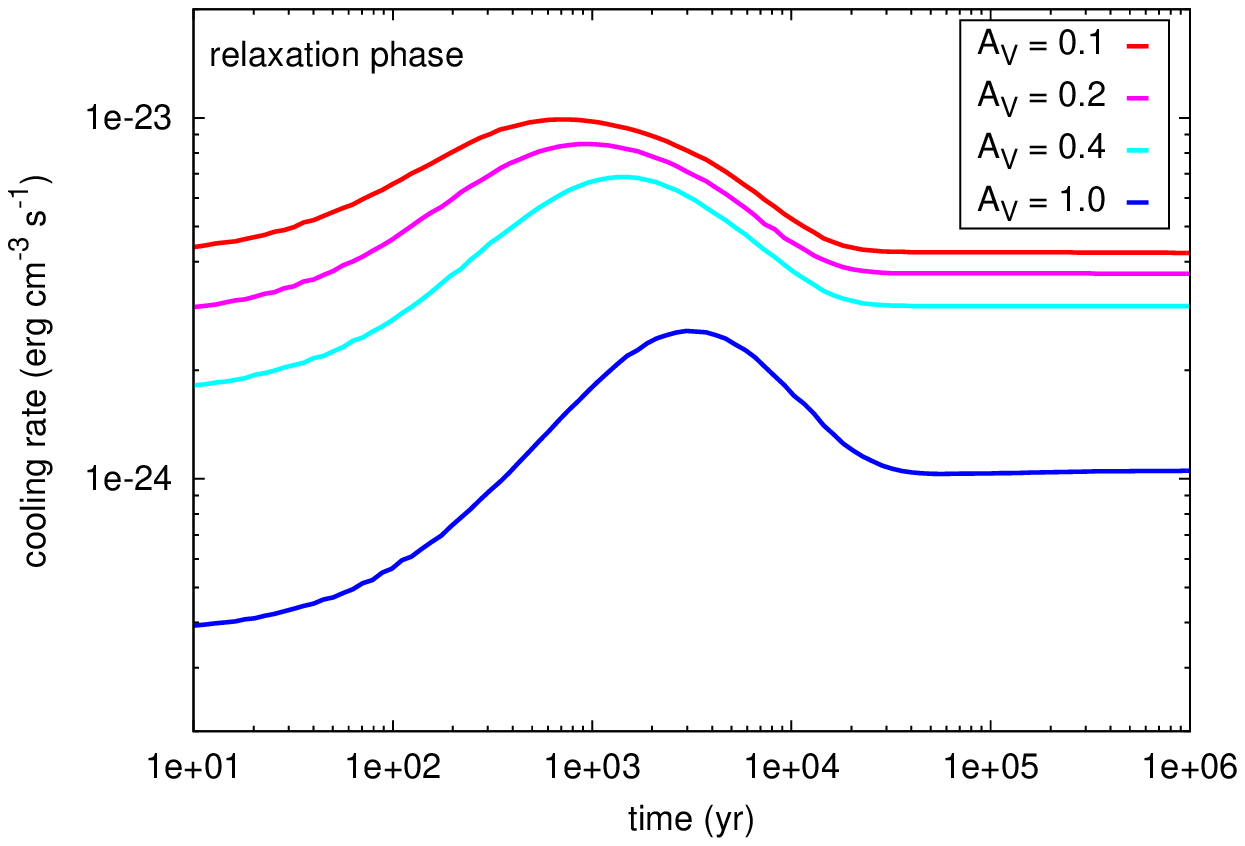}
\caption{{\it Top panel}: \Cp\ relative abundance as a function of the distance from the 
vortex axis during the active stage of the TDR model. {\it Bottom panel}: cooling rate 
associated with the $^2P_{3/2} \rightarrow ^2P_{1/2}$ transition of \Cp\ as a function of 
time during the relaxation stage of the TDR model. The model is computed for $\dens = 100$ 
\cc, $a= 10^{-11}$ s$^{-1}$, and several values of the visible extinction $A_V$: $0.1$ 
(red), $0.2$ (magenta), $0.4$ (cyan), and $1.0$ (blue) mag.}
\label{Fig-Cp-TDR}
\end{center}
\end{figure}

To illustrate the impact of turbulent dissipation on the excitation of the [CII] line, 
we display in Fig. \ref{Fig-Cp-TDR} the relative abundance of \Cp\ as a function of the 
position in an active vortex of the TDR model (top panel) and the cooling rate due to 
the [CII] line during the relaxation period (bottom panel) for a point located at the 
$r_0$ radius of a vortex (see Sect. \ref{Sec-TDR} and Paper I). We note that in this 
figure, the time axis is expressed in log units to allow the visualisation of the [CII] 
line cooling evolution. The origin at $t= 10$ yr is somewhat arbitrary and is meant 
to allow a display of the [CII] emission close to the origin of the relaxation phase.

In an active vortex, the ignition of reaction \ref{Reac-CHp-text} opens a fast chemical 
loop starting from \Cp\ that successively increases the production of \CHdp, \CHtp, and 
of neutral carbon (see Fig. \ref{Fig-Network-CHp} of Appendix \ref{Append-Networks}). 
The bottleneck of this loop being the slow photoionisation of C, the abundance of \Cp\ 
decreases across the active vortex (top panel of Fig. \ref{Fig-Cp-TDR}). The intensity 
of the [CII] line emitted during the active phase ($I_{\rm [CII]}$ at the shortest time 
(or before) in the bottom panel of Fig. \ref{Fig-Cp-TDR}), is therefore smaller than that 
emitted by the ambient medium ($I_{\rm [CII]}$ at times larger than a few $10^4$ yr in 
the bottom panel of Fig. \ref{Fig-Cp-TDR}). Once the active stage stops, the abundance 
of \Cp\ quickly increases up to its initial value, while the thermal relaxation lasts 
much longer. The outcome is a period during which the abundance of \Cp\ is high and the
temperature still high enough to enhance the [CII] emission. This period lasts for more 
than $\sim 10^4$ yr while the active phase lasts only for a few hundred to one thousand 
years (see Sect. \ref{Sec-oxy-TDR}). Turbulent dissipation is therefore responsible for 
an enhancement of [CII] emission above that of the ambient gas.

The resulting [CII] intensities computed with TDR models combined with a PDR-like reconstruction
of the line of sight are shown in the bottom panel of Fig. \ref{Fig-Cp-FIR}. For small values 
of the rate-of-strain, we find that the predictions of the model are in good agreement 
with the observations, that a PDR-like shielding function along the line of sight accounts 
for the slower increase in \Cp\ emission at large $I_{\rm FIR}$ and that the scatter of 
observational data could be explained by two indistinguishable scenarios: either an
increase in the fragmentation of the gas or variations of \epso, $E_{\tau_V}$, or $a$.

The [CII] line emission of diffuse and transluscent gas is therefore enhanced by the 
heating of the gas due to turbulent dissipation compared to PDR predictions.

\section{Discussion} \label{Sect-discussion}

\subsection{The turbulent dissipation scenario inferred from chemistry}

\begin{figure*}[!ht]
\begin{center}
\includegraphics[width=8.0cm,angle=-0]{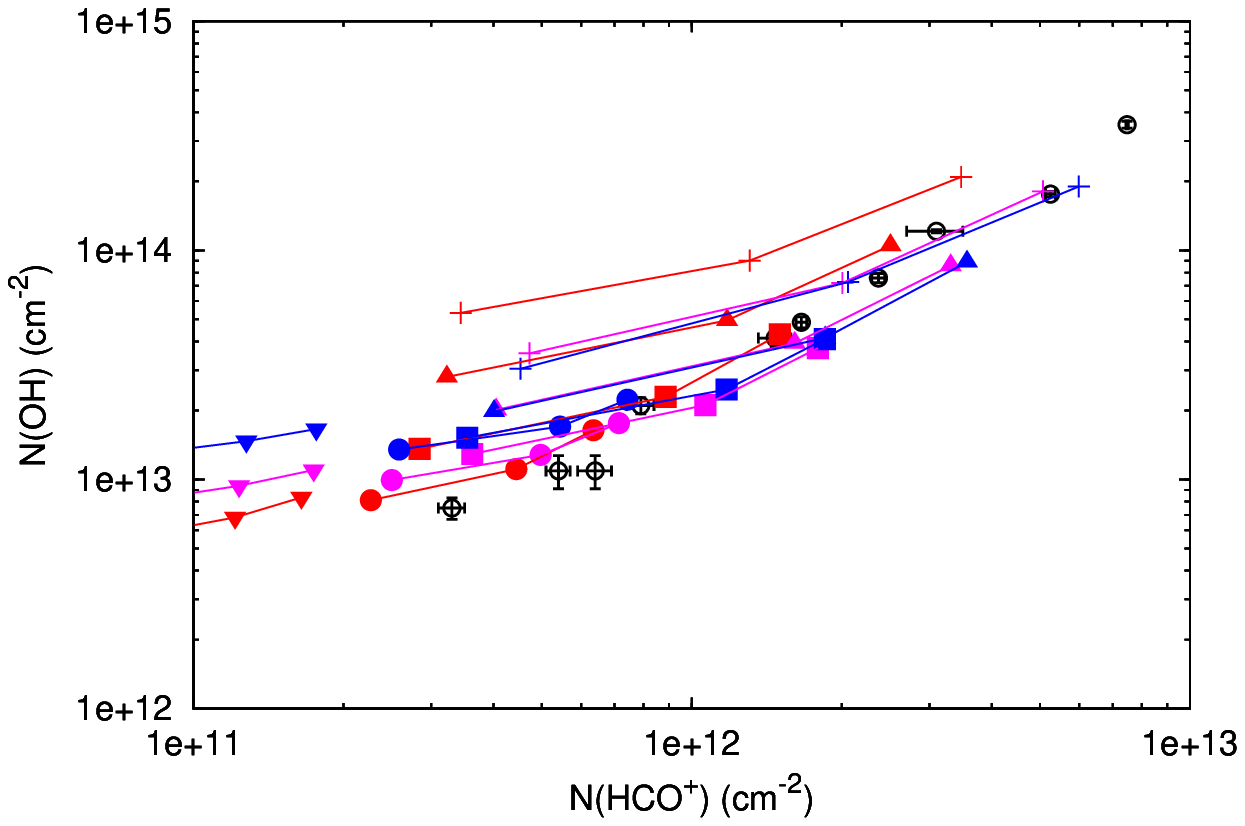}
\includegraphics[width=8.0cm,angle=-0]{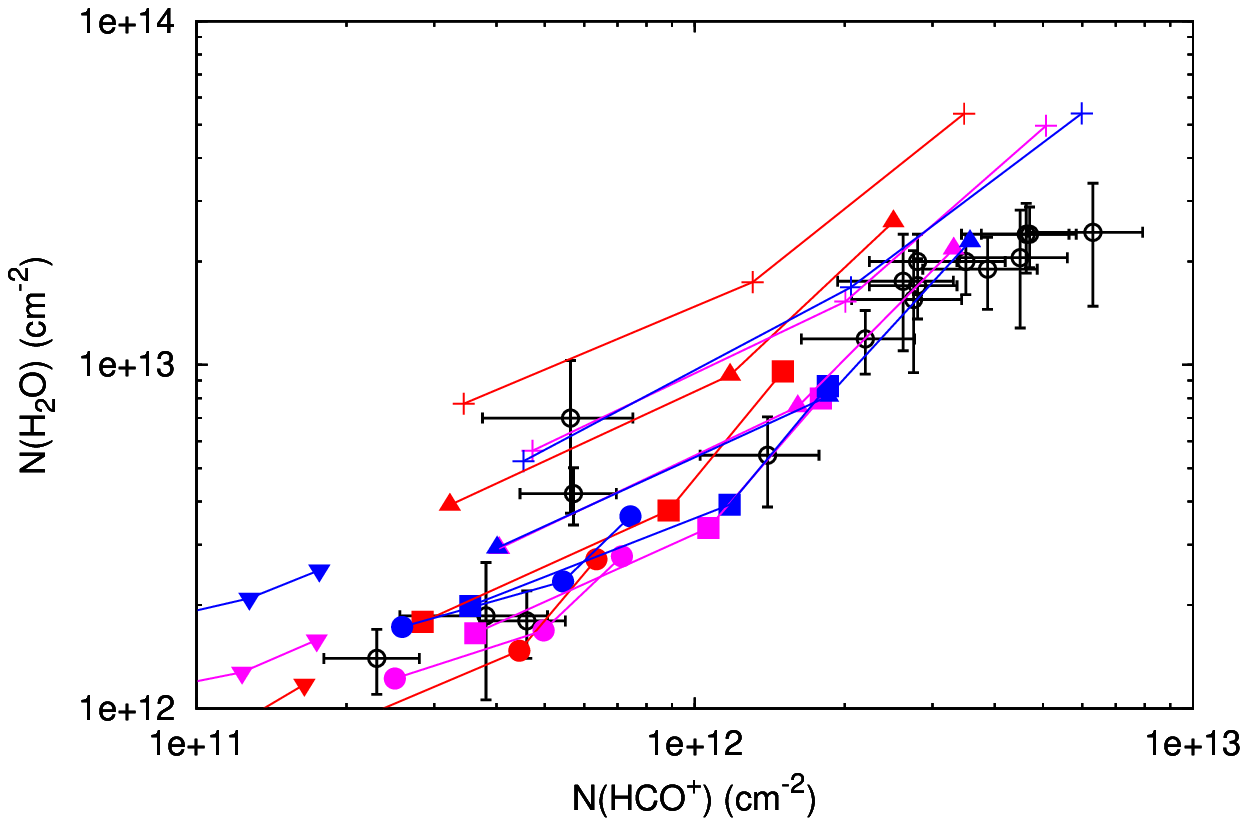}
\includegraphics[width=8.0cm,angle=-0]{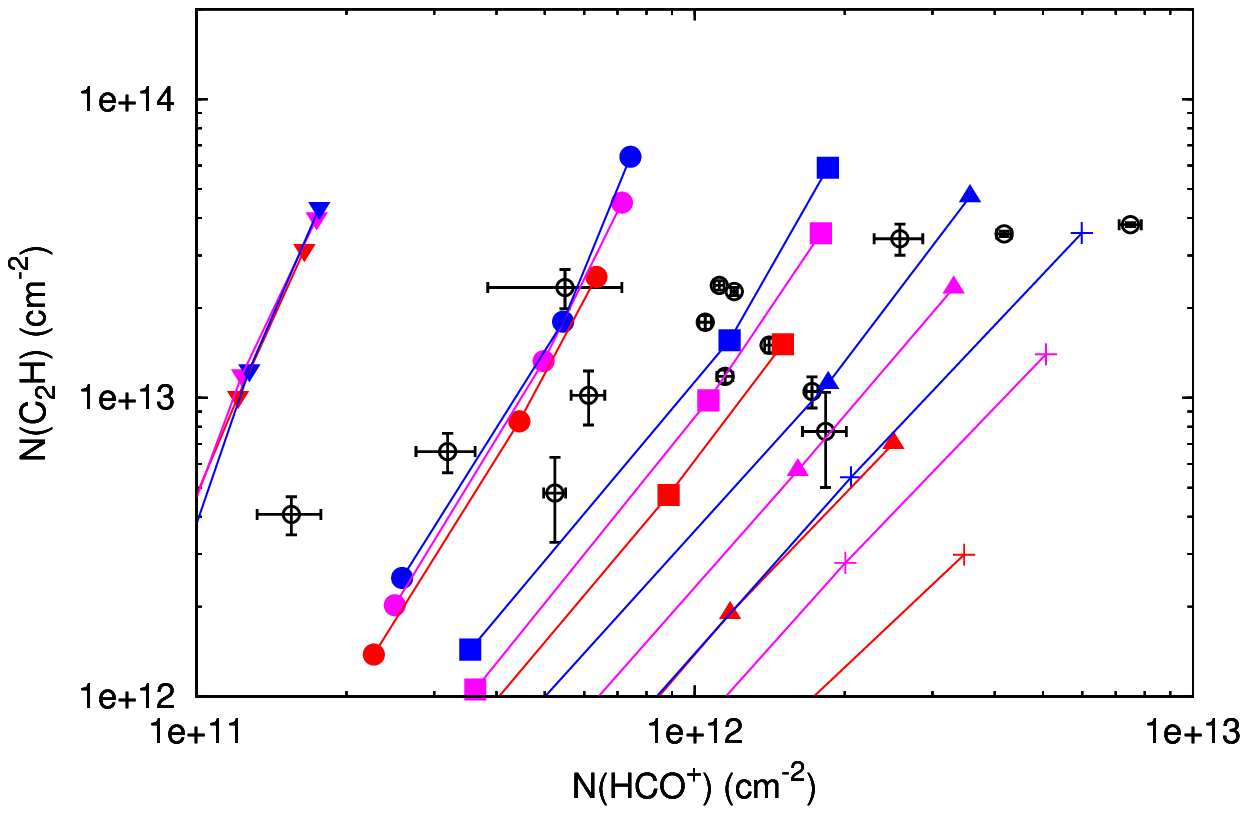}
\includegraphics[width=8.0cm,angle=-0]{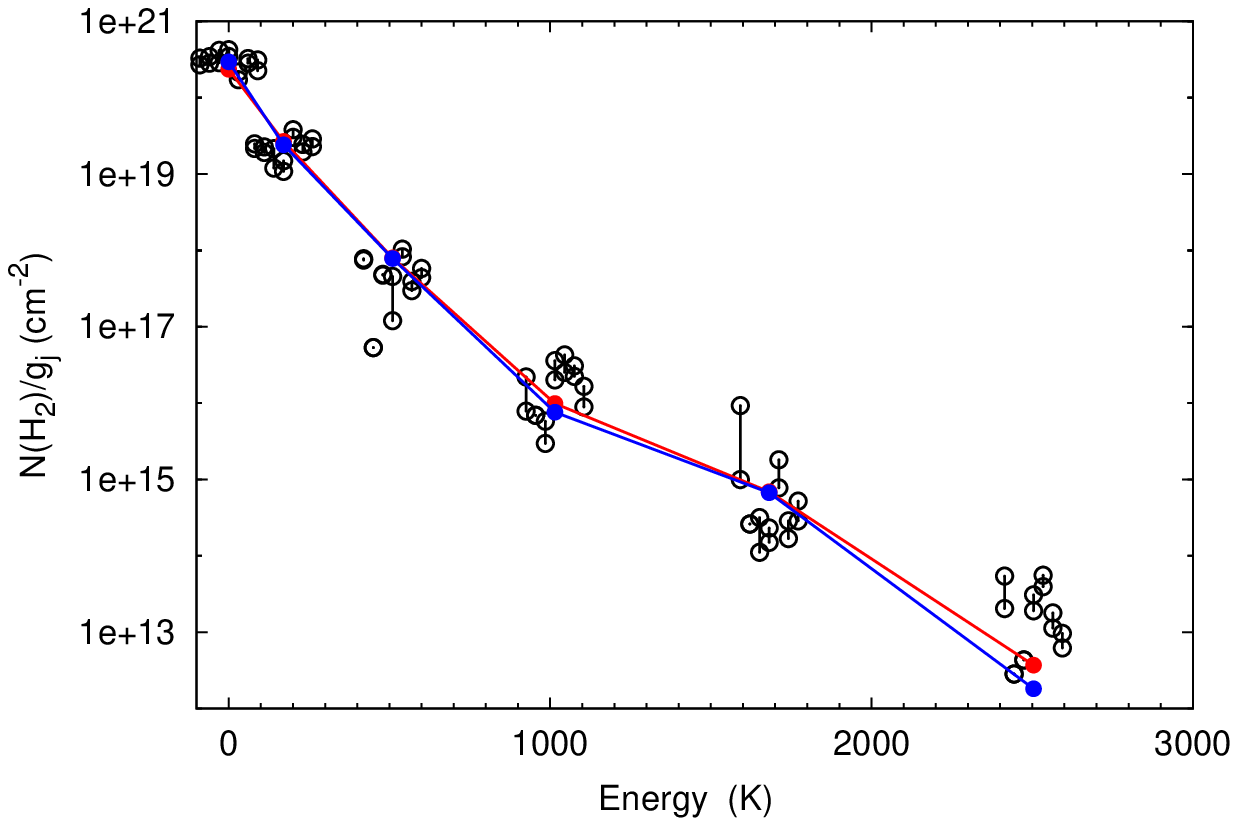}
\caption{Observations compared to TDR model predictions. Data (open circles): column
densities of \HCOp\ and OH from \citet{Lucas1996} ({\it top left panel}), \HdO\ from 
\citet{Flagey2013} ({\it top right panel}), \CtH\ from \citet{Lucas2000} ({\it bottom left panel}), 
and the excited levels of \HH\ from \citet{Gry2002} and \citet{Lacour2005} ({\it bottom right 
panel}). For clarity, in the bottom right panel the points for a given level are slightly 
shifted on the abscissa. Models (filled symbols): computed for $A_V=0.2$ (red), 0.4 
(magenta), and 0.6 (blue) mag and $\dens = 20$ (crosses), 30 (triangles), 50 (squares), 
100 (circles), and 300 (upside-down triangles) \cc. Except for the bottom right panel, 
$a$ increases along each curve from $10^{-11}$ to $10^{-10}$ s$^{-1}$ from right to left. 
In the bottom right panel, the models are shown for $A_V=0.1$ mag (red) and $0.4$ mag (blue) 
and single values of the rate-of-strain $a = 3\times 10^{-11}$ s$^{-1}$ and the 
density $\dens = 100$ \cc.}
\label{Fig-add-TDR}
\end{center}
\end{figure*}

The previous sections have shown that, despite unavoidable degeneracies, the comparison of 
the TDR model results with the observations identifies optimal ranges of three of the free 
parameters related to turbulent dissipation: $\epso \lesssim 10^{-23}$ \eccs\ from the 
largest abundances of \CHp, $\utM \lesssim 3.5$ \kms\ from the \SHp/\CHp\ abundance ratio, 
and $ 10^{-13} \lesssim E_{\tau_V} \lesssim 10^{-12}$ erg cm$^{-1}$ from the CO/\HCOp\ 
abundance ratio. We found in Paper I that the chemical properties of the diffuse ISM favour 
small rates-of-strain with the optimal range  $10^{-11} \lesssim a \lesssim 10^{-10}$ 
s$^{-1}$ that has been used in the present paper. Without fine tuning, these parameters
reproduce the column densities of three additional molecular species OH, \HdO, and \CtH, 
and their link with those of \HCOp. They also reproduce the populations of the first five 
rotational levels of \HH\ observed in the local diffuse matter through UV absorption 
spectroscopy \citep{Gry2002,Lacour2005} (Fig. \ref{Fig-add-TDR}). These results are in 
good agreement with those of Paper I, even though the chemical network has been updated 
(see Appendix \ref{Append-Chemistry}) because the local abundances of all these hydrides 
both in UV-driven and turbulence-driven chemistries are determined by only a handful of 
chemical reactions (see Appendix \ref{Append-Networks}), the rates of which are well 
constrained by laboratory experiments and theoretical calculations of reaction dynamics.

The outcome of the whole study is therefore an ensemble of four parameter ranges, related 
to turbulent dissipation, that have been determined independently from different molecular 
species, or pairs of species, and that satisfactorily reproduce the abundances of many other 
species observed in the diffuse ISM. We discuss in Sect. \ref{Sect-consist} how these values 
compare with the known properties of turbulence in the diffuse ISM.

\subsection{Influence of the cosmic ray ionisation rate on the TDR model predictions} \label{Sect-zeta}

\begin{figure}[!ht]
\begin{center}
\includegraphics[width=8.5cm,angle=-0]{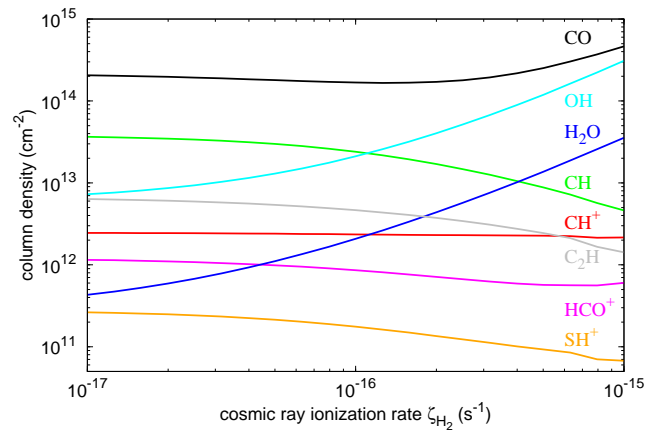}
\caption{Column densities of CH, \CHp, \SHp, OH, \HdO, \HCOp, and CO predicted with the
TDR model for a cloud of total column density $\NHt = 1.8 \times 10^{21} \cq$ as functions 
of the \HH\ ionisation rate $\zetaHH$. All the other parameters of the code are set to 
their standard values (see main text).}
\label{Fig-zeta}
\end{center}
\end{figure}


Since the low-energy cosmic ray fluxes cannot be observed from Earth (partly because of the
interplanetary magnetic field) the current estimations of the cosmic ray ionisation rate 
of atomic or molecular hydrogen\footnote{$\zetaHH = 2.3 / 1.7 \times \zetaH$ \citep{
Glassgold1974}.} ($\zetaH$ and $\zetaHH$, respectively) mostly rely on the observations 
of molecular tracers. Such methods are, however, uncertain because they strongly depend on 
the chemical models used to interpret the observations. Studies of dense clouds with OH, 
\HCOp, and HD indicate that $\zetaH$ is in the range $1 - 5$ $\times 10^{-17}$ s$^{-1}$ 
with an average of $\sim 3 \times 10^{-17}$ s$^{-1}$ \citep{van-der-Tak2000}. Similarly, 
early studies of diffuse clouds with the same species have led to $\zetaH \sim 5 \times 
10^{-17}$ s$^{-1}$ in the direction of $\zeta$ Oph, $o$ Per, $\chi$ Oph, $\zeta$ Per, and 
$\xi$ Per \citep{van-Dishoeck1986,Federman1996a}. However, the recent detections of \Htp\ 
\citep{McCall2004,Indriolo2007,Indriolo2012}, \OHp, and \HdOp\ \citep{Gerin2010a,Neufeld2010a,
Indriolo2014} in the diffuse gas now point towards values considerably larger. Indeed the 
comparison of these data with chemical models solely driven by the UV radiation and the 
cosmic ray particles shows that $\zetaH \sim 2 \times 10^{-16}$ s$^{-1}$ on average (e.g. 
\citealt{Hollenbach2012}).


In this paper, we have therefore systematically run the TDR model for three different values 
of $\zetaHH$: $3 \times 10^{-17}$, $10^{-16}$, and $3 \times 10^{-16}$ s$^{-1}$. In addition, 
and in order to further explore the impact of this parameter, 20 models were run covering the 
range $10^{-17} \leqslant \zetaHH \leqslant 10^{-15}$ s$^{-1}$; all the other parameters were 
set to their standard values (see Sect. \ref{Sect-explo-param}). The standard model, $\zetaHH = 
10^{-16}$ s$^{-1}$, corresponds to the best-fit scenario of all the observational data (column 
densities and column density ratios) with the TDR model. We discuss here the sensitivity 
of our results to $\zetaHH$.

Figure \ref{Fig-zeta} shows that most of the species considered in this work are not sensitive 
to the ionisation rate of \HH. The column densities of \CHp, \SHp, \HCOp, CO, and \CtH\ vary 
by less than a factor of four when $\zetaHH$ increases by two orders of magnitude. It follows 
that $\zetaHH$ has a limited impact on the determination of the properties of turbulent 
dissipation (\epso, \utM, and $E_{\tau_V}$, see Sects. \ref{Sect-CHp}, \ref{Sect-SHp}, and 
\ref{Sect-CO}) and thus weakly affects the conclusions of this paper. In contrast, Fig. 
\ref{Fig-zeta} shows that CH, OH, and \HdO\ are particularly sensitive to $\zetaHH$. The 
column density of CH decreases by a factor of 10 and those of OH and \HdO\ increase by two 
orders of magnitude when $\zetaHH$ increases from $10^{-17}$ to $10^{-15}$ s$^{-1}$ because 
$\zetaHH$ boosts both the productions of H and \Hp\ which respectively destroy CH (via CH + 
H $\rightarrow$ C + \HH) and initiate the oxygen hydrogenation chain (via O + \Hp\ 
$\rightarrow$ \Op\ + H).

In Fig. \ref{Fig-add-TDR} we find that the column densities of OH and \HdO\ predicted with 
the TDR model are in good agreement with the observations for $\zetaHH = 10^{-16}$ s$^{-1}$. 
It is worth noting that the column density of CH predicted for this value ($N({\rm CH}) \sim 
2 \times 10^{13}$ \cq) is in agreement with those observed in the local diffuse medium along 
lines of sight with total column density $\NHt = 1.8 \times 10^{21} \cq$ \citep{Sheffer2008}.

These results differ from those of \citet{Indriolo2012}, \citet{Hollenbach2012}, and 
\citet{Indriolo2014} who found $\zetaHH \sim 3 \times 10^{-16}$ s$^{-1}$ with PDR-type 
models and the recent observations of \Htp, \OHp, and \HdOp. The main reason for this 
difference can be found in Fig. \ref{Fig-Triangles} which shows that the ambient medium 
accounts for only about 50 \% of the column densities of OH and \HdO\ in the framework 
of the TDR model. Indeed, it implies that the ionisation rate inferred from the TDR model 
is about two times smaller than that inferred from PDR-type models. In addition, as shown 
by \citet{Wolfire2008} and \citet{Hollenbach2012}, the predictions of PDR-type models on 
carbon and oxygen chemistry depend on the details of the modeling of the polycyclic aromatic 
hydrocarbons (PAH) in the computation of the chemical reaction rates and in particular 
their size, geometry and abundances. The difference between our optimal value of $\zetaHH$ 
and that deduced by \citet{Hollenbach2012} could therefore be due to the fact that we model 
PAHs as spherical particles of radius 6.4 \AA\ (or equivalently $\sim 120$ carbon atoms) 
and relative abundance $\sim 4 \times 10^{-7}$ while they consider planar PAHs containing 
100 carbon atoms and with a relative abundance of $2 \times 10^{-7}$.

\subsection{Consistency with known properties of turbulence in the diffuse ISM} \label{Sect-consist}

The upper limit of \epso\ was derived from the largest values of $N(\CHp)/\NHt$ detected in 
the diffuse medium towards the inner Galaxy. The parameter \epso\ is the injection rate of 
turbulent energy at the large scales, $\rho \sigma_l^3/l$, where $\rho$ is the gas volumic 
mass and $\sigma_l$ the rms velocity dispersion at scale $l$. This quantity is difficult to 
determine observationally in the atomic gas because it involves a dynamical time $l/\sigma_l$ 
and therefore a sizescale $l$, while the available data are only Gaussian velocity components. 
As discussed in Sect. \ref{sec-data}, this value has been estimated in the atomic medium in 
only one region of the SN \citep{Miville-Deschenes2003}, so no statistics are available. In 
diffuse molecular gas traced by CO(1-0), large statistics are now available (\citealt{Hennebelle2012})
and the upper limit on \epso\ is of the order of the largest values of $\epso_{obs}$ measured 
at scales between 10 pc and 50 pc that can be considered as relevant large scales for turbulence 
in the diffuse medium. Moreover, these values are themselves averages measured at each of these 
large scales. The fact that \CHp\ is not detected with average relative abundances larger than 
$10^{-7}$ confirms previous estimations of the maximum turbulent energy density contained in 
the diffuse ISM. The lower limit of $N(\CHp) /\NHt$ cannot be exploited because it comes from 
the sensitivity limit of the observations. The observed large scatter by a factor of $\sim 10$ 
of the \CHp\ abundance (Fig. 4) could be interpreted by the fact that the formation of \CHp\ 
is highly intermittent: the \CHp\ abundance traces subsets of the CNM where the turbulent 
transfer rate is above average, hence driving an above-average dissipation rate. The above-average 
dissipation rate needs nonetheless to heat the gas to temperatures that compare with the 
endothermicity threshold of the \CHp\ formation.
 
The upper limit on the ion-neutral drift velocity rules out values larger than 3.5 \kms. In 
the model, this velocity drift is set by the maximum tangential velocity of the vortex, which 
is itself linked to the velocity dispersion of the ambient turbulence, i.e. a property of 
turbulence discussed in Sect. \ref{Sec-TDR}. It is noteworthy that this value of 3.5 \kms\ 
is close to the large-scale velocity dispersion of the CNM seen in the HI line \citep{
Miville-Deschenes2003,Haud2007} and is also representative of the internal velocity dispersion 
of the diffuse molecular gas at scales of 10 -- 50 pc. The parameters relevant to the ambient 
turbulence, inferred independently from the \CHp\ and \SHp\ abundances, are therefore broadly 
consistent with our knowledge of turbulence in the diffuse medium.

\subsection{Internal coherence}
We now turn to the parameters controlling the dynamics of the dissipative regions and 
compare the different associated timescales. The optimal range of rates-of-strain given 
above corresponds to timescales $a^{-1}$ varying between 300 yr and 3000 yr. The independent 
determination of the burst lifetime, $\tau_V$, is in the range  100 -- 1000 yr for the 
reference vortex (see Sect. \ref{Sect-CO}). Finally, the turnover timescale of a vortex, 
$2\pi r_0/\utM$ is of the order of $10^3$ yr for $r_0=100$ au and $\utM=3$ \kms. All these 
timescales are therefore commensurate, confirming the picture of a highly non-equilibrium 
chemistry driven by turbulent dissipation bursts over timescales as short as 100 yr and size 
scales of $\sim 100$ au. It is interesting to note that the small values of $a$, favoured 
by the chemistry, lead to timescales longer than the vortex lifetime, i.e. the action of the 
rate-of-strain lasts at least as long as the vortex lifetime.


\subsection{Limitations of the TDR model}

Some of the limitations of the TDR model have already been discussed in Paper I. In particular, 
the use of only one type of dissipation burst was mentioned, while dissipation is known to be 
distributed among a large number of bursts of different strengths \citep{Mouri2009}. The warm 
chemistry being driven by endoenergetic reactions with exponential thresholds, the abundances 
and abundance ratios of several species greatly depend on the strength of the burst as long as 
the corresponding thresholds are not overcome (see Sect. \ref{Sect-SHp}). A distribution of 
bursts could thus have an important impact. Implementing such a distribution in the TDR code 
would require defining additional parameters that cannot be constrained with the current 
observations of the diffuse ISM.


Our model is idealised, but the results rely on two robust properties of gas-phase 
interstellar chemistry: the large differences of endothermicities and the orders of 
magnitude differences between relaxation times due to the various destruction mechanisms 
(photodissociation, recombination, collisions with H and \HH). Hence, the results of the 
TDR model presented here do not depend on the details of the treatment of the regions of
dissipation. Its originality  relies in the exploitation of the fact that the chemical 
timescales in the diffuse medium are commensurate with the smallest dynamical times of 
turbulence, at which its huge reservoir of energy is dissipated.

\subsection{Turbulent dissipation in vortices and shear layers or C-shocks?}

The problem raised by the high abundances of \CHp\ in the diffuse medium was recognised 
several decades ago. Additional gas heating mechanisms were invoked and modelled to 
reproduce the observed column densities: gas heating behind shocks associated with HII 
regions expansion \citep{Elitzur1978}, ion-neutral friction in shocks with magnetic 
precursors, i.e. C-shocks \citep{Draine1986a,Pineau-des-Forets1986}, or in Alfv\'en 
waves that also involve the streaming of \Cp\ relative to \HH\ \citep{Federman1996}.

To favour one scenario over the others from the sole observations of chemical abundances 
and line profiles is difficult. For instance \CHp\ could be explained by one C-shock 
with a velocity of 10-12 \kms\ leading to a predicted velocity shift between \CHp\ and 
CH of 2 \kms\ at most, i.e. if the shock propagates along the line of sight \citep{
Flower1998c}. Yet, it is more likely that several low velocity shock waves will be present 
along the line of sight, with different orientations and velocities. \citet{Gredel2002} 
proposed such a model with $\sim 20$ C-shocks of velocities of $\sim 8$ \kms\ per $A_V$, 
leading to a velocity shift between CH and \CHp\ smaller than 1 \kms.

However, it is now understood that shocks are not the only structures where dissipation is 
concentrated in compressible turbulence. In transonic hydrodynamic turbulence, \citet{
Porter2002} found that the interaction of shocks generates vorticity, resulting in only 
a small fraction of the kinetic energy left in compressional modes. In supersonic MHD 
turbulence, the dynamic alignment of velocity and magnetic fields tends to Alfv\'enize 
turbulence, a process at the origin of the low ratio of the energy density in compressional 
to solenoidal modes \citep{Kritsuk2010}. Similar conclusions were found in other simulations 
of MHD turbulence where incompressible modes carry more than 70 \% of the total kinetic energy 
\citep{Vestuto2003}. It is for these reasons that we chose to investigate the chemical
signatures of turbulent dissipation in vortices.
%

\section{Conclusions} \label{Sect-conclusion}

The vast diversity of the chemical formation and destruction pathways of atoms and molecules,  
and of their timescales in particular, provide unique clues to the non-equilibrium physical 
processes driving the thermal and chemical evolution of the diffuse ISM that have never been 
investigated before. In this paper we have presented a comprehensive study of the chemical 
composition of diffuse matter in the framework of UV-driven and turbulence-driven chemistries. 
The analysis focuses on 16 different atomic and molecular species observed towards several 
hundred sightlines sampling the SN and the inner Galaxy diffuse medium. 

We have distinguished the species that are the most sensitive to turbulent dissipation 
(the abundances of \CHp\ and \SHp, and the populations of the $J=3,4,5$ rotational levels 
of \HH) from those that are not and trace the interaction between the UV radiation field 
and the interstellar matter (H, \HH, HF, and the $J=0,1,2$ excited levels of \HH). The 
column densities of CO and \HCOp and the intensity of the [CII] line emission are also 
found to be sensitive to turbulent dissipation. 

The chemical differences among all these species are so pronounced that robust results on 
the turbulent cascade in the diffuse ISM can be inferred from the observations of specific 
pairs of species. In the framework of the TDR model, we find that (1) the \CHp\ abundance 
is proportional to the average turbulent dissipation rate, thus to the average energy transfer 
rate in the turbulent cascade; (2) the \SHp/\CHp\ abundance ratio sets an upper limit to the 
ion-neutral velocity drift in the active dissipation regions because of the large difference 
of endothermicities involved in the formation of these two species; and (3) the \HCOp/CO 
abundance ratio sets the lifetime of the active dissipation bursts because of the large
difference of destruction timescales of these two species. We find that the increase in \CHp\ 
abundance in the inner Galaxy compared to the value derived in the local ISM could be due to 
an increase in the turbulent injection rate towards the inner Galaxy. This set of independently 
determined properties provides a coherent scenario in which the regions of active dissipation 
of turbulence in the diffuse ISM are small-scale (a few $100$ au) short-lived ($\sim 100 - 
1000$ yr) bursts in which dissipation is mainly due to ion-neutral decoupling ($\sim 3$ \kms)
rather than viscous friction. These results are in good agreement with independent determinations 
of the turbulence properties in the diffuse ISM and with the fundamental properties of turbulence 
known from laboratory experiments.

We also show for the first time that turbulent dissipation plays a crucial role in the 
formation of diffuse interstellar CO and stress the influence of the fragmentation of 
the gas on the observed correlation between CO and molecular hydrogen: large column 
densities of CO at low $N(\HH)$ may result from the accumulation on the line of sight 
of fragments, each weakly shielded from the UV radiation field. Finally, we show that 
turbulent dissipation enhances the [CII] line emission in transluscent environments by 
a factor of $\sim 2$ compared to the value obtained if the gas heating was solely driven 
by UV photons.

These results shed a new light on the interpretation of atomic and molecular data and 
provide templates for the analysis of future observations of diffuse environments in Galactic 
and extragalactic sources.

\begin{acknowledgements}

We are most grateful to James G. Ingalls for providing the data used in Fig. \ref{Fig-Cp-FIR}.

\end{acknowledgements}

\bibliographystyle{aa} 
\bibliography{mybib}


\appendix

\section{Treatment of the chemistry} \label{Append-Chemistry}

\begin{table}[!ht]
\begin{center}
\caption{Elemental abundances $[{\rm X}]=n({\rm X})/\dens$ used in the TDR model as 
measured in the local interstellar medium \citep{Anders1989,Savage1996,Sofia2001,
Snow2007}.}
\begin{tabular}{l c}
\hline
$[{\rm He}]$ &  $1.00 \times 10^{-01}$ \\
$[{\rm C }]$ &  $1.38 \times 10^{-04}$ \\
$[{\rm O }]$ &  $3.02 \times 10^{-04}$ \\
$[{\rm N }]$ &  $7.94 \times 10^{-05}$ \\
$[{\rm S }]$ &  $1.86 \times 10^{-05}$ \\
$[{\rm F }]$ &  $1.80 \times 10^{-08}$ \\
$[{\rm Cl}]$ &  $1.00 \times 10^{-07}$ \\
$[{\rm Fe}]$ &  $1.50 \times 10^{-08}$ \\
\hline
\end{tabular}
\label{Tab-Ele}
\end{center}
\end{table}

In our previous papers \citep{Godard2009, Godard2010, Godard2012}, the chemical network
implemented in the TDR model was built as a combination of the chemical network of the
Meudon PDR code \citep{Le-Petit2006} and those extracted from the main online databases 
for astrochemistry\footnote{available at \url{http://www.udfa.net/}, \url{http://kida.obs.u-bordeaux1.fr},
and \url{http://www.physics.ohio-state.edu/~eric/research.html}.}, UMIST \citep{Woodall2007,McElroy2013}, 
OSU \citep{Hassel2010}, and KIDA \citep{Wakelam2012}. To improve the usefulness of the 
model we break here from the previous approach and choose to adopt the chemical network of 
the Meudon PDR code (available at \url{http://pdr.obspm.fr/PDRcode.html}) except for a few 
special cases presented below.

\subsection{Chlorine and fluorine chemistries}

By default the chemical network of the Meudon PDR code contains 133 gas-phase atomic
and molecular compounds $-$ including hydrogen-, helium-, carbon-, nitrogen-, oxygen-,
sulfur-, silicon-, and iron-bearing species $-$ which interact with one another through 
2631 gas-phase reactions. In order to obtain TDR predictions concerning the abundances 
of HF, HCl, and HCl$^+$, all recently detected in the diffuse ISM with Herschel/HIFI 
\citep{Neufeld2010,Sonnentrucker2010,Monje2013,De-Luca2012}, the network has been expanded 
to 14 chlorine- and fluorine- bearing species: Cl, HCl, CCl, Cl$^+$, HCl$^+$, H$_2$Cl$^+$, 
CCl$^+$, H$_2$CCl$^+$, F, HF, F$^+$, HF$^+$, H$_2$F$^+$, and CF$^+$. This additional 
network includes the 38 chemical reactions given in the recent review of experimental 
and theoretical studies performed by \citet{Neufeld2009}. Their rates are those of 
\citet{Neufeld2009} except for the photodissociation of HF, HCl, and HCl$^+$ and the
photoionisation of Cl, the rates of which are taken from the UMIST database \citep{McElroy2013}
and the calculations of \citet{van-Dishoeck1988} and \citet{van-Dishoeck2006}. This
additional network also includes the 38 ion-neutral and dissociative recombination
reactions given by \citet{Anicich2003} and the OSU database for astrochemistry 
\citep{Hassel2010}.

\subsection{Photoreactions}

Since the amounts of matter of prototypical diffuse and transluscent environments 
\citep{Snow2006} are insufficient to entirely absorb the mean interstellar UV radiation 
field, both photoionisation and photodissociation are known to play a major role 
in the destruction of atoms and molecules in the diffuse ISM. In the Meudon PDR code
\citep{Le-Petit2006} both processes are treated self-consistently by solving the UV 
radiative transfer as a function of position in a plan parallel slab $-$ including the 
absorption by molecular lines and the absorption and scattering by interstellar dust $-$ 
and by computing at each point the photodestruction rates from the integration of the 
cross sections over the specific intensity of the UV radiation.

Such a detailed approach cannot, however, be applied to the TDR code since the radiative 
transfer is modelled with a single parameter, the extinction at visible wavelength $A_V$, 
assumed to be constant over the entire sightline (standard case, see Sect \ref{Sec-build-los}). 
The photo-reaction rates are thus computed as
\begin{equation} \label{eq-photo}
k_\gamma = \gamma {\rm exp}(-\beta A_V) \quad {\rm s}^{-1},
\end{equation}
where $\gamma$ and $\beta$ are constants taken from \citet{van-Dishoeck1988} and
\citet{van-Dishoeck2006} who performed fits of the photodestruction rates over the range
$A_V = 0-3$ mag, assuming a slab of gas with constant density illuminated on one side and 
where the extinction is only the result of the absorption and scattering by 0.1 $\mu$m 
large interstellar grains.

While this expression is reliable for the photodissociation occurring though continuous 
absorption, it is however too simplistic to correctly describe the photodissociation by 
line absorption processes which strongly depends on the self-shielding, i.e. the column 
density of the molecule from the edge of the cloud up to $A_V$. We therefore apply Eq.
\ref{eq-photo} to all molecules except for \HH\ and CO whose photodissociation is known
to occur through line processes \citep{Lee1996,Draine1996,Visser2009}. For these two 
species we adopt the photo-reaction rates computed with the Meudon PDR code assuming a 
slab of gas with constant density illuminated on one side. The details of these results 
and their implementation in the TDR model are described in Appendix \ref{Append-H2-CO}.

\subsection{Excitation of \HH}

In addition to the chemical state of the gas the TDR model also follows the time-dependent 
evolution of the level populations of \HH\ in the three different phases, the ambient medium, 
the active stage, and the relaxation stage. While the code was built to treat up to 318
different levels of molecular hydrogen, we only consider here the first 18 rovibrational
levels (with an energy above the ground state smaller than $10^4$ K) in order to reduce 
the computation time.

Since the TDR model differs from PDR models by the absence of radiative transfer, neither 
the UV radiative pumping nor the near- and mid-infrared absorption and induced emission 
are taken into account. The populations of the excited levels of \HH\ thus results from
the combined effects of collisional excitation with H, He, and \HH\ \citep{Flower1998,
Flower1998a,Flower1998b,Le-Bourlot1999} and spontaneous radiative decay. This latter 
process is included in the gas cooling function assuming that all the rovibrational 
lines of \HH\ are optically thin, a hypothesis that holds in the diffuse ISM as long as 
$N_{\rm H} \lesssim 4 \times 10^{24}$ cm$^{-2}$ for a gas velocity dispersion of 1 \kms.

\subsection{State-to-state chemical rates}

During the past decades, the advances of the cross molecular beam experiments, the flowing 
afterglow apparatus, the ion trapping techniques, and the theoretical studies of chemical 
reaction dynamics (see, e.g. the reviews of the field by \citealt{Casavecchia2000,
Levine2005}) have led to measurements and calculations of state-specific reaction rates for 
several neutral-neutral and ion-neutral reactions (e.g. \citealt{Zanchet2013}). In particular
the internal energy of \HH\ has been proven to systematically increase the reactivity of 
highly endothermic reactions \citep{Hierl1997,Sultanov2005,Zellner1981}. To take this 
process into account, we have implemented in the TDR model a state-specific description
of the reaction rates of
\begin{equation} \label{Reac-CHp}
{\rm C}^+ + \HH(\upsilon,J) \rightarrow \CHp + {\rm H} \quad (\Delta E/k = 4640 K) \quad {\rm and}
\end{equation}
\begin{equation} \label{Reac-SHp}
{\rm S}^+ + \HH(\upsilon,J) \rightarrow \SHp + {\rm H} \quad (\Delta E/k = 9860 K).
\end{equation}
For the first reaction, we follow the approach of \citet{Agundez2010} and adopt the 
state-specific rate constants of \citet{Gerlich1987} for $\HH(\upsilon=0,J=0...7)$ and
the Langevin collision rate for higher energy levels \citep{Hierl1997}. Since there is
no information concerning the state-to-state rate constants of reaction \ref{Reac-SHp},
we assume
\begin{equation}
k = \left \{
    \begin{array}{l l}
       1.1 \times 10^{-10} {\rm exp}\left[-\frac{9860-E(\upsilon,J)}{T_{\rm eff}} \right] {\rm cm}^3{\rm s}^{-1} & {\rm if}\,\, E(\upsilon,J) \leqslant 9860 {\rm K} \\
       1.6 \times 10^{-9} {\rm cm}^3{\rm s}^{-1} & {\rm otherwise}
    \end{array}
    \right .,
\end{equation}
where $E(\upsilon,J)$ is the energy (expressed in K) of the level $\upsilon,J$ of \HH.

Comparing the predictions of the model with those obtained without including 
state-to-state chemistry \citep{Godard2009}, we find that these mechanisms have a 
little impact on the production of \CHp\ and \SHp\ in the diffuse gas.

\subsection{Electron / ion recombinations on grains and PAHs}

Electron transfers between ions and very small grains or polycyclic aromatic hydrocarbons 
(PAHs) are efficient in decreasing the ionisation state of atoms and molecules \citep{
Lepp1988}. When implemented in astrochemical models, such processes are found 
to compete with (or even dominate) the radiative and dielectronic recombinations of several 
ions such as H$^+$ or \Cp\ \citep{Bakes1998,Welty2003,Wolfire2008,Hollenbach2012} and thus 
to have a strong impact on the hydrogenation chains and the charge balance of carbon- and 
oxygen-bearing species (see Fig. \ref{Fig-Network-CO}).

To take these mechanisms into account we have implemented in the TDR code a treatment of 
the charge of PAHs and very small grains using the formalism described by \citet{Bakes1994} 
and \citet{Weingartner2001a}. Large grains are neglected. Similarly to our treatment of 
photoreactions, the photoionisation rates of dust particles are modelled using Eq. 
\ref{eq-photo} with coefficients inferred from the computations of the Meudon PDR code. 
The electronic attachment and ion recombinations 
on, respectively, the positively and negatively charged dust particles are modelled using the
prescription of \citet{Draine1987} and taking into account the electron affinity, the escape 
length, and the ionisation potentials of the colliders in the computation of the rates
\citep{Weingartner2001,Weingartner2001a}.

Both PAHs and very small grains are described as spherical particles of radius 6.4 \AA\ 
and 13 \AA, respectively. Their abundances in the diffuse ISM are parametrised by the size 
distribution function of interstellar dust and the dust mass fraction with respect to 
the gas phase. In this paper we consider a dust-to-gas mass ratio of 0.01 and assume that 
the PAHs carry 4.6 \% of the total dust mass \citep{Draine2007}. The size distribution of
PAHs is modelled as a log-normal function centered at 6.4 \AA\ \citep{Compiegne2011}. The
size distribution of very small grains is modelled as a power law with an index of -3.5.
All these considerations lead to $n({\rm PAHs}) / \dens = 4.2 \times 10^{-7}$ and 
$n({\rm VSGs}) / \dens = 2.5 \times 10^{-8}$.

\section{Photodissociation rate of \HH\ and CO: self-shielding} \label{Append-H2-CO}

\begin{figure}[!ht]
\begin{center}
\includegraphics[width=8.5cm,angle=-0]{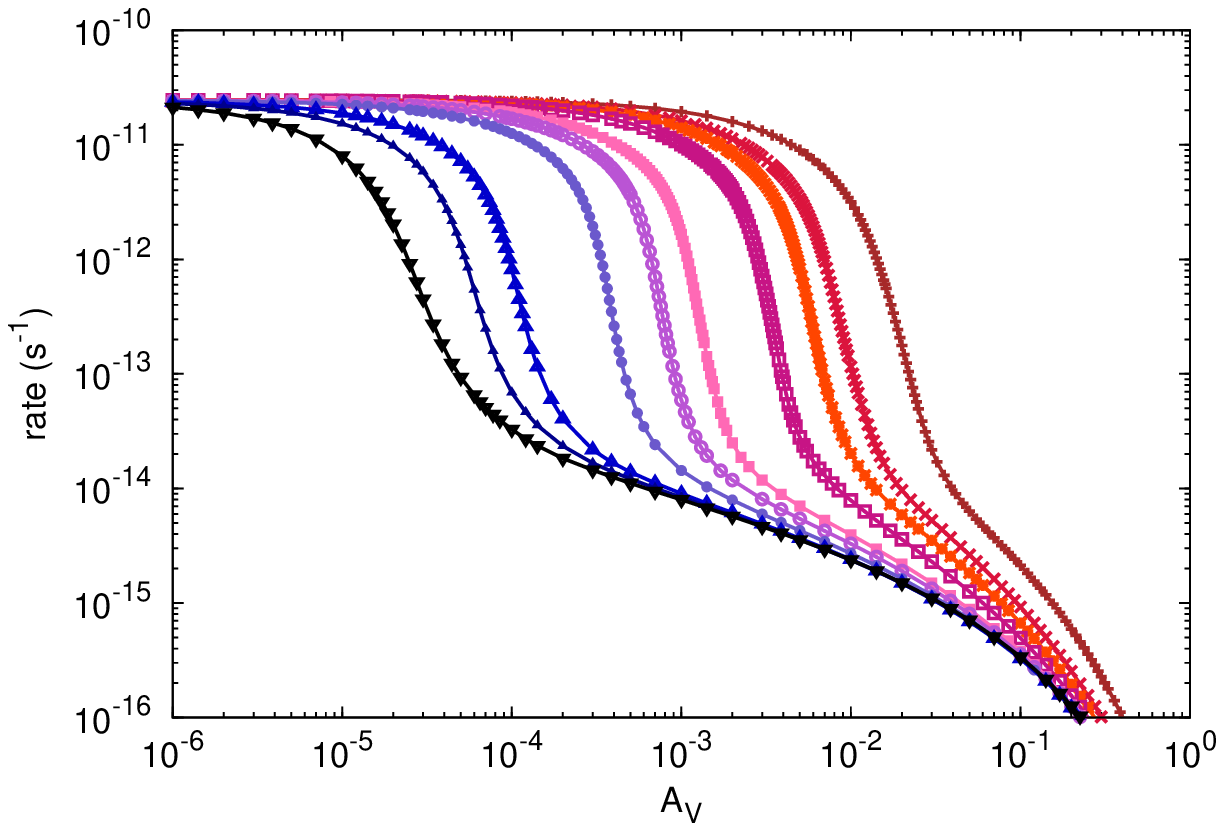}
\caption{\HH\ photodissociation rate as a function of the extinction computed with the 
Meudon PDR code applied to PDRs with different densities illuminated on one side: from 
right to left $\dens=10$, $30$, $50$, $100$, $300$, $500$, $1000$, $3000$, $5000$, and 
$10000$ \cc, successively.}
\label{Fig-Photo-H2}
\end{center}
\end{figure}

\begin{figure}[!ht]
\begin{center}
\includegraphics[width=8.5cm,angle=-0]{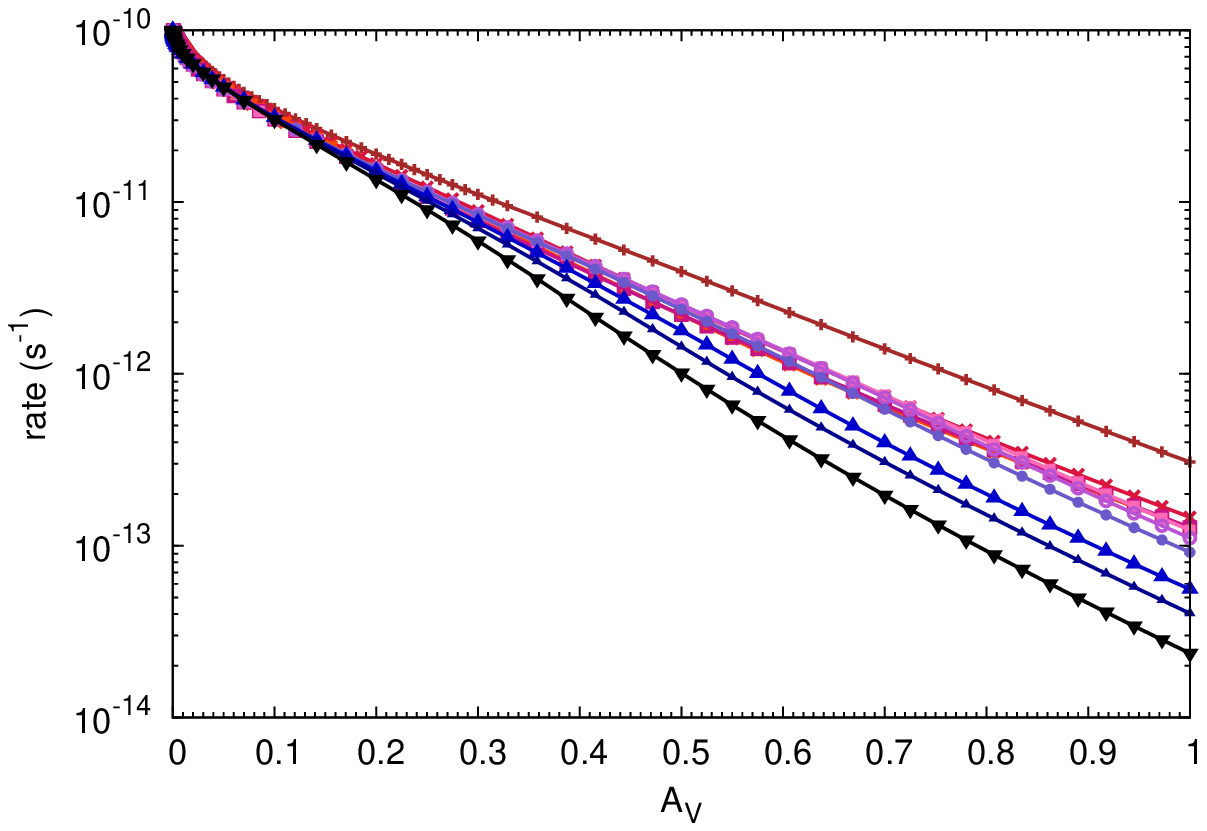}
\caption{CO photodissociation rate as a function of the extinction computed with the 
Meudon PDR code applied to PDRs with different densities illuminated on one side: from 
right to left $\dens=10$, $30$, $50$, $100$, $300$, $500$, $1000$, $3000$, $5000$, and 
$10000$ \cc, successively.}
\label{Fig-Photo-CO}
\end{center}
\end{figure}

\begin{figure}[!ht]
\begin{center}
\includegraphics[width=8.5cm,angle=-0]{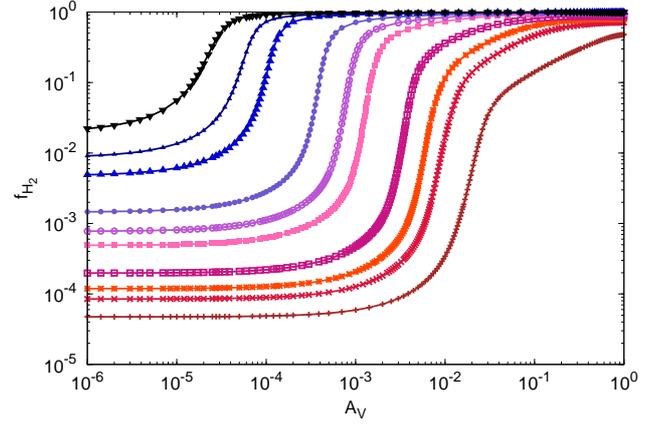}
\caption{Molecular fraction defined as $f_{\HH} = 2 n(\HH) / \dens$ as a function of $A_V$
computed with the Meudon PDR code for several densities: from right to left $\dens=10$, $30$,
$50$, $100$, $300$, $500$, $1000$, $3000$, $5000$, and $10000$ \cc, successively.}
\label{Fig-TrHH2}
\end{center}
\end{figure}

In the standard case of the TDR code, the radiative transfer is modelled with a single 
parameter $A_V$, the extinction of the radiation field in the visible photometric band 
($\lambda \sim 0.551$ $\mu$m). Such a prescription is, however, too simplistic to correctly 
describe the photodissociations of \HH\ and CO which occur through line processes and 
therefore strongly depend on the self-shielding \citep{van-Dishoeck1986,van-Dishoeck1988a}. 
Because these photodissociations drive 
the chemical transitions from H to \HH\ and from C to CO, it is essential to take into 
account their dependence on $A_V$.

The distinctive dependences of the photodissociation rates of \HH\ and CO on $A_V$, 
$N(\HH)$, and N(CO) have been studied analytically and numerically by \citet{Draine1996} 
and \citet{Lee1996}. In their approaches \citet{Draine1996} consider a directional UV 
radiation field propagating perpendicularly to an infinite plan-parallel slab of gas, 
while \citet{Lee1996} apply the approximation proposed by \citet{Federman1979} to solve 
the self-shielding of \HH\ and CO. We have decided to part from these previous studies 
and to compute the photodissociation rates of \HH\ and CO assuming an isotropic radiation 
field and taking into account the impact of the line broadening on the self-shielding.

The Meudon PDR code was therefore run along a grid of models defined by the following 
range of parameters : $10^1 \leqslant \dens \leqslant 10^4$ \cc, $0.5 \leqslant \chi 
\leqslant 10$ (in Mathis's units), and $3 \times 10^{-17} \leqslant \zetaHH \leqslant 3 
\times 10^{-16}$ s$^{-1}$. The code was set to solve the exact transfer within the 
electronic lines of \HH\ and CO assuming a Doppler broadening of 3.5 \kms. While the
code is designed to compute self-consistently the formation rate of \HH\ on grain surfaces
using both the Eley–Rideal and Langmuir-Hinshelwood mechanisms \citep{Le-Bourlot2012,Bron2014}, 
we switched off these processes and set the \HH\ formation rate 
to its observed value of $3 \times 10^{-17} (T/100 {\rm K})^{1/2} \dens n({\rm H})$ cm$^{-3}$ 
s$^{-1}$ where $n({\rm H})$ is the density of atomic hydrogen. The obtained photodissociation 
rates of \HH\ and CO and the predicted molecular fractions of the gas are shown in Figs. 
\ref{Fig-Photo-H2}, \ref{Fig-Photo-CO}, and \ref{Fig-TrHH2} as functions of $A_V$ and for 
the subset of models $\chi = 1$ and $\zetaHH= 3 \times 10^{-16}$ s$^{-1}$.

The obtained photodissociation rates of \HH\ differ from those derived by \citet{Draine1996} 
by less than a factor of three for $N(\HH) \lesssim 2 \times 10^{21}$ \cq\ and by more than an order 
of magnitude for $N(\HH) \gtrsim 3 \times 10^{21}$ \cq. The former difference comes from the effect
of the line broadening on the self-shielding while the latter is ascribed to the different
prescriptions (directional or isotropic) used for the radiative transfer. These interpretations
are confirmed with the comparison of our results with those computed by \citet{Lee1996}. In
this case the photodissociation rates of \HH\ are found to differ by less than a factor of 
three over the whole range of $N(\HH)$ depending on the values of the density and the cosmic 
ray ionisation rate $\zetaHH$.

\section{PDR and TDR chemical networks} \label{Append-Networks}

\begin{figure*}[!hb]
\begin{center}
\includegraphics[width=15cm,angle=-0]{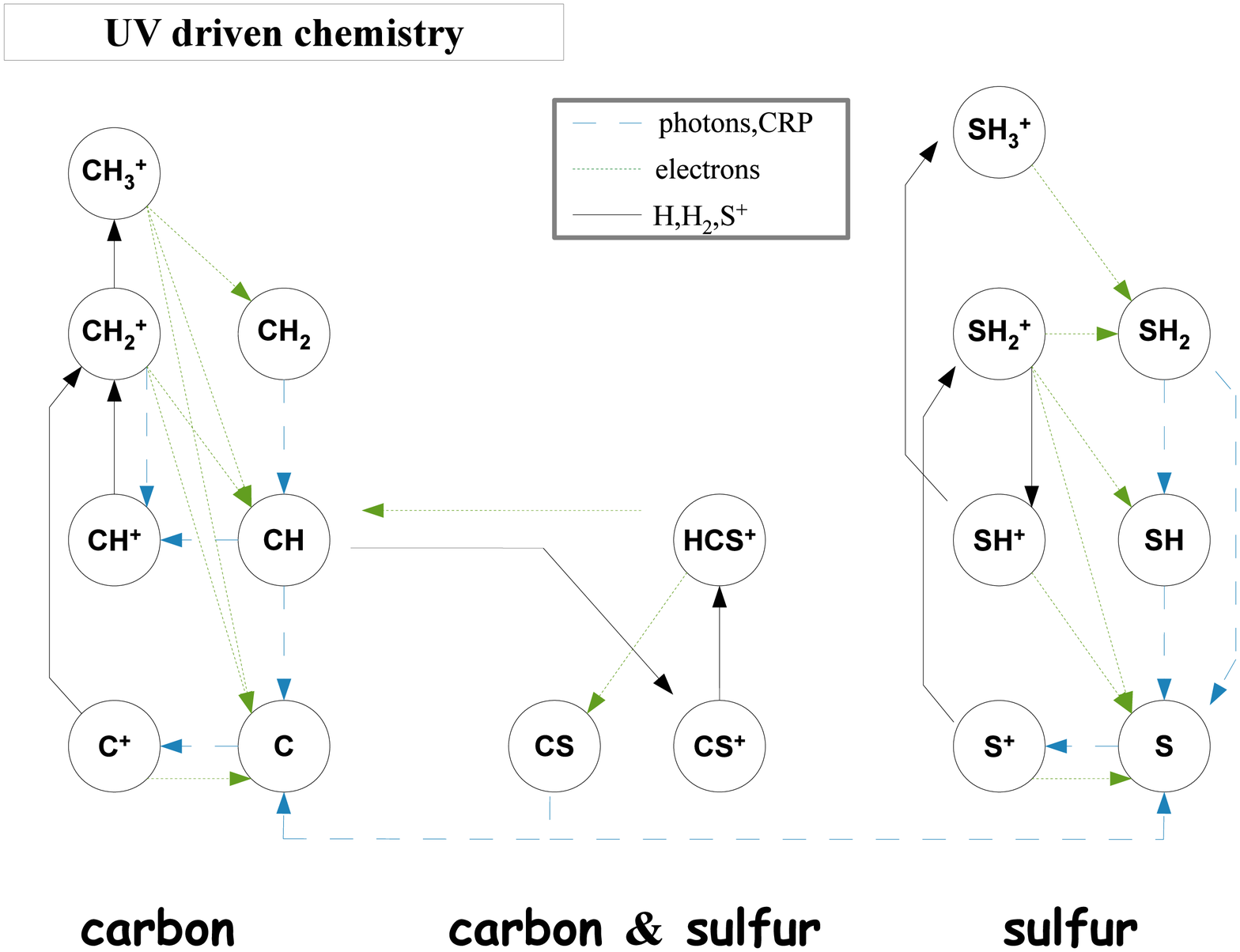}
\includegraphics[width=15cm,angle=-0]{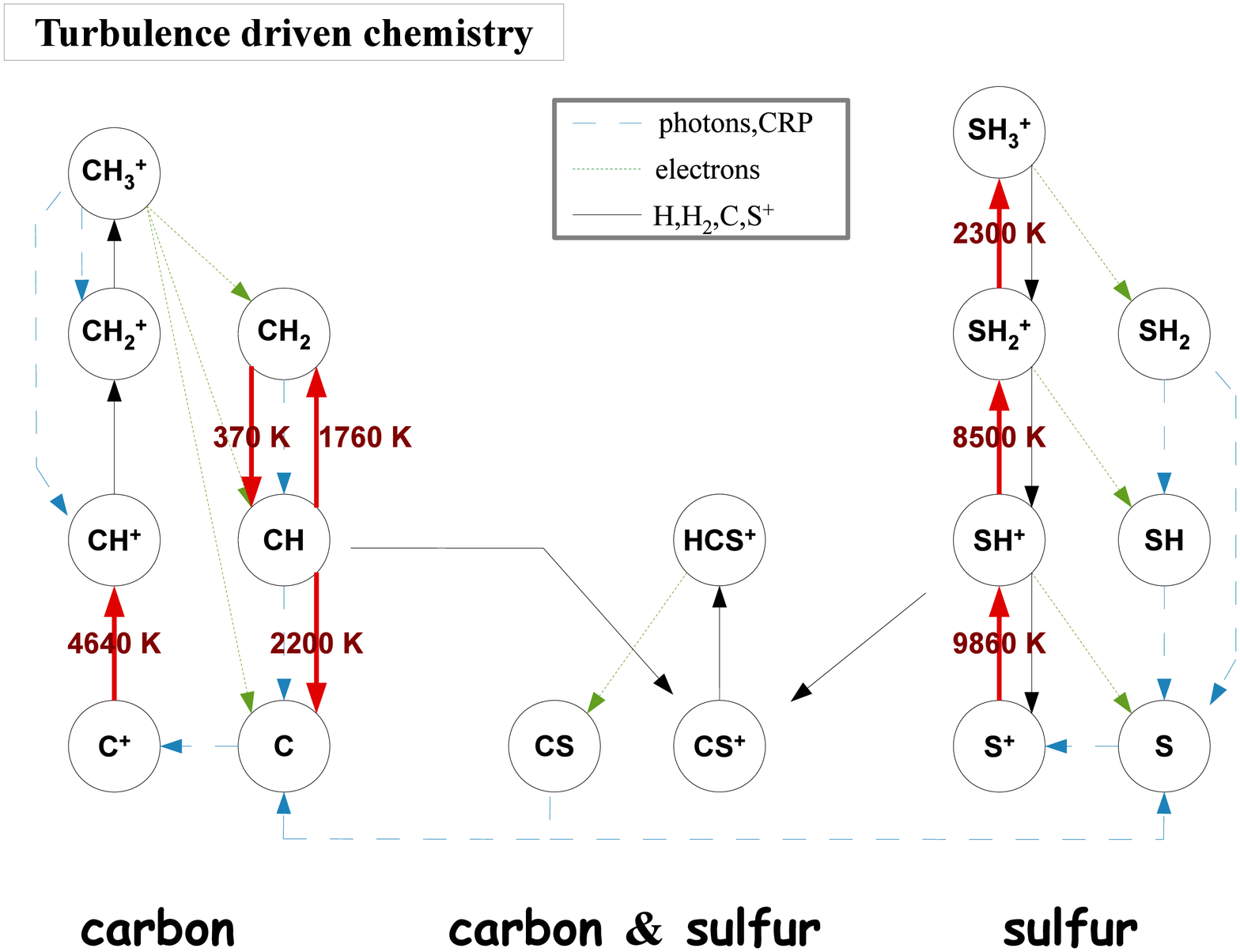}
\caption{Carbon and sulfur chemical networks driven by the UV radiation field (top panel) 
and the turbulent dissipation (bottom panel) in the diffuse interstellar medium assuming 
$\dens = 50$ \cc\ and $A_V$ = 0.4 mag. This figure is simplified: for each species, only the 
processes that together contribute to more than 70 percent of the total destruction and 
formation rates are displayed. The endothermicities and energy barriers of endoenergic 
reactions are indicated in red.}
\label{Fig-Network-CHp}
\end{center}
\end{figure*}

\begin{figure*}[!hb]
\begin{center}
\includegraphics[width=15cm,angle=-0]{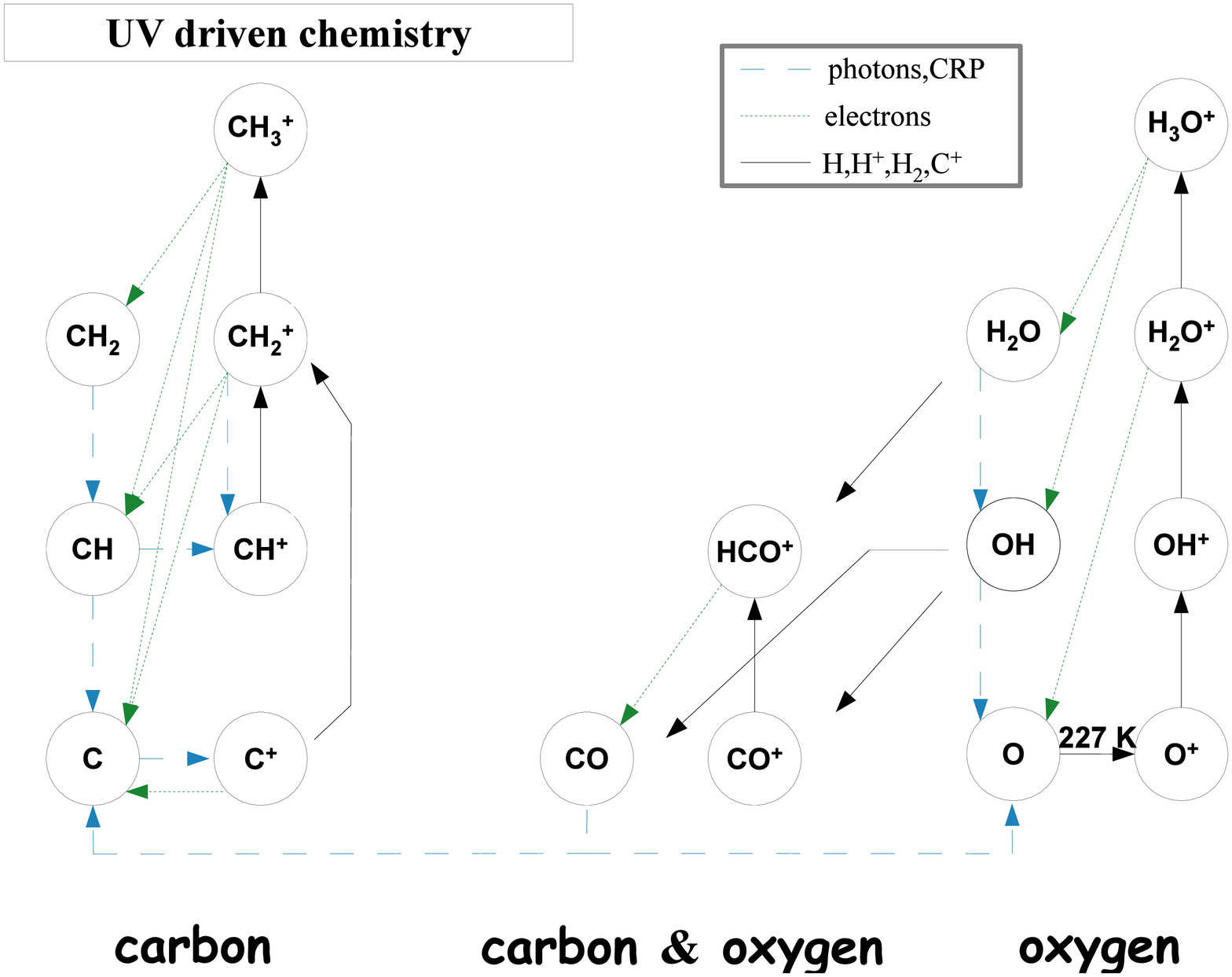}
\includegraphics[width=15cm,angle=-0]{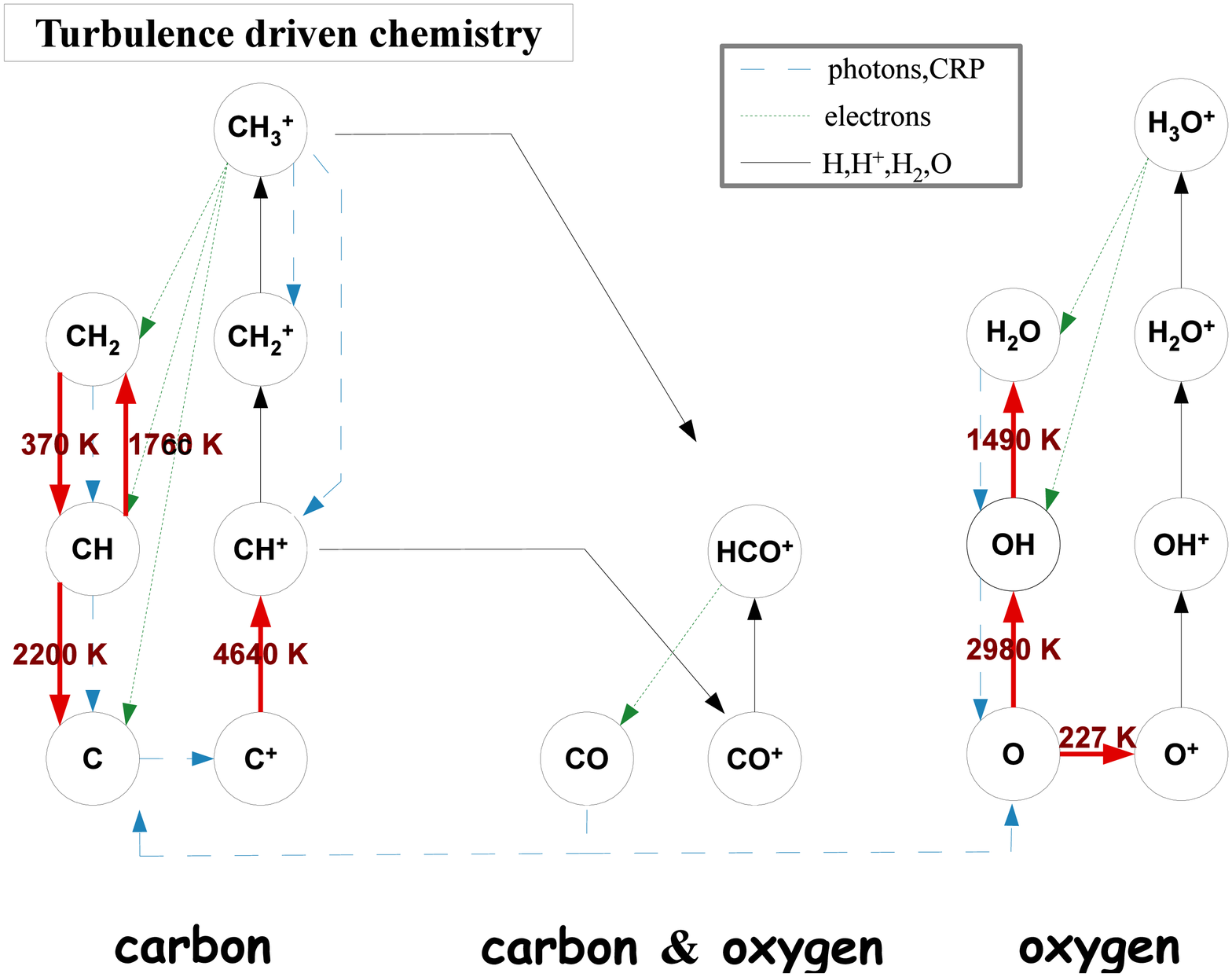}
\caption{Carbon and oxygen chemical networks driven by the UV radiation field (top panel) 
and the turbulent dissipation (bottom panel) in the diffuse interstellar medium assuming 
$\dens = 50$ \cc\ and $A_V$ = 0.4 mag. This figure is simplified: for each species, only the 
processes that together contribute to more than 70 percent of the total destruction and 
formation rates are displayed. The endothermicities and energy barriers of endoenergic 
reactions are indicated in red.}
\label{Fig-Network-CO}
\end{center}
\end{figure*}

The chemical compositions of photodissociation regions and turbulent dissipation regions 
are driven by very different chemical patterns. This occurs because the heating of PDRs 
is dominated by the interaction of UV photons with interstellar matter while that of
TDRs is controlled by turbulent dissipation, but also because the state-of-the-art PDR 
models systematically neglect the gas dynamics and thus solve the at-equilibrium chemical
state, while TDR models focus on the non-equilibrium effects, i.e. on the coupling between 
the dynamics and the chemistry as a function of time. To illustrate these differences we show
in Figs. \ref{Fig-Network-CHp} and \ref{Fig-Network-CO} the main production and destruction 
routes of 27 species including the carbon, sulfur, and oxygen hydrogenation chains as given 
by the PDR and TDR models. In both cases we show the production pathways obtained locally 
assuming a diffuse molecular gas of density $\dens=50$ \cc, a shielding $A_V = 0.4$ mag, 
and a cosmic ray ionisation rate $\zetaHH = 10^{-16}$ s$^{-1}$. For the TDR model, the 
results are extracted at the $r_0$ radius (see Paper I) of an active vortex set by the 
following parameters: $\utM = 3.5$ \kms and $a=10^{-11}$ s$^{-1}$. All these figures are 
simplified: for each species shown, only the processes that together contribute to more 
than 70 percent of the total destruction and formation rates are shown. 

\end{document}